\newif\ifdraft
\draftfalse

\newif\ifcameraready
\camerareadytrue

\newif\iffullversion
\fullversiontrue

\PassOptionsToPackage{table}{xcolor}
\PassOptionsToPackage{hyphens}{url}

\iffullversion
\documentclass[onecolumn, 11pt]{article}
\else
\documentclass[sigconf]{acmart}

\copyrightyear{2018}
\acmYear{2018} 
\setcopyright{iw3c2w3}
\acmConference[WWW 2018]{The 2018 Web Conference}{April 23--27, 2018}{Lyon, France}
\acmBooktitle{WWW 2018: The 2018 Web Conference, April 23--27, 2018, Lyon, France}
\acmPrice{}
\acmDOI{10.1145/3178876.3186099}
\acmISBN{978-1-4503-5639-8/18/04}

\fi

\iffullversion
\usepackage{amsfonts, amsmath}
\usepackage[margin=1in]{geometry}
\usepackage{url}
\usepackage[sort]{cite}
\fi

\usepackage{xspace}
\usepackage[capitalize,noabbrev]{cleveref}
\usepackage{algorithm}
\usepackage[noend]{algpseudocode}
\usepackage{subcaption}
\usepackage{tikz}
\usetikzlibrary{shapes}
\usepackage{balance}

\definecolor{Gray}{gray}{0.85}
\newcolumntype{a}{>{\columncolor{Gray}}c}

\usepackage[\ifdraft draft \else final \fi]{changes}
\definechangesauthor[color=red]{S}
\definechangesauthor[color=blue]{A}
\definechangesauthor[color=teal]{P}
\definechangesauthor[color=orange]{R}
\definechangesauthor[color=brown]{N}
\setauthormarkup{}

\renewcommand{\paragraph}[1]{\vspace{2mm}\noindent \textbf{#1.\ }}

\newcommand{\nw}{Ripple network\xspace}
\newcommand{\link}{credit link\xspace}
\newcommand{\links}{credit links\xspace}
\newcommand{\acc}{wallet\xspace}
\newcommand{\accs}{wallets\xspace}

\newcommand{\Accs}{Wallets\xspace}
\newcommand{\tx}{transaction\xspace}
\newcommand{\txs}{transactions\xspace}
\newcommand{\Txs}{Transactions\xspace}
\newcommand{\sdr}{sender\xspace}
\newcommand{\rcv}{receiver\xspace}
\newcommand{\xrp}{XRP\xspace}

\newcommand{\users}{users\xspace}
\newcommand{\sndr}{sender\xspace}

\newcommand{\uid}{\ensuremath{u}\xspace}

\newcommand{\credit}{credit\xspace}
\newcommand{\rflag}{\textsf{no\_ripple}\xspace}
\newcommand{\dflag}{\textsf{defaultRipple}\xspace}

\newcommand{\resil}{resilience\xspace}

\newcommand{\resilfactor}{{rsl-factor}\xspace}

\newcommand{\set}[1]{\{#1\}\xspace}
\newcommand{\wid}{\ensuremath{w}\xspace}

\newcommand{\g}[1]{\textsf{gr-#1}\xspace}
\newcommand{\gall}{\textsf{gr-\{13-17\}}\xspace}

\newcommand{\mm}[1]{\textsf{mm-#1}\xspace}

\newcommand{\s}[1]{\textsf{skeleton-#1}\xspace}

\newcommand{\txdataset}{\textsf{tx-\{13-17\}}\xspace}
\newcommand{\gwdataset}{\textsf{gw-17}\xspace}
\newcommand{\prung}[1]{\textsf{pruned-g-#1}\xspace}

\newcommand{\liqg}[1]{\textsf{liq-g-#1}\xspace}

\newcommand{\snodes}{\textsf{100-deg}\xspace}
\newcommand{\tnodes}{\textsf{100-ftx}\xspace}

\newcommand{\stuckacc}[2]{\textsf{stuck-wallets-#1-#2}\xspace}

\newcommand{\EE}{\ensuremath{\mathbb{E}}\xspace}
\newcommand{\VV}{\ensuremath{\mathbb{V}}\xspace}

\newcommand{\payroutes}{PayRoutes\xspace}
\newcommand{\dr}{DividendRippler\xspace}

\ifdraft
\newcommand{\TODOA}[1]{\added[id=A]{{\bf ToDo: #1}}}
\newcommand{\TODOS}[1]{\added[id=S]{{\bf ToDo: #1}}}
\newcommand{\TODOP}[1]{\added[id=P]{{\bf ToDo: #1}}}
\newcommand{\TODOR}[1]{\added[id=R]{{\bf ToDo: #1}}}
\newcommand{\TODON}[1]{\added[id=N]{{\bf ToDo: #1}}}
\newcommand{\DISCUSS}[1]{\color{red} \textbf{#1} \color{black}}
\else
\newcommand{\TODOA}[1]{\added[id=A]{}}
\newcommand{\TODOS}[1]{\added[id=S]{}}
\newcommand{\TODOP}[1]{\added[id=P]{}}
\newcommand{\TODOR}[1]{\added[id=R]{}}
\newcommand{\TODON}[1]{\added[id=N]{}}
\newcommand{\DISCUSS}[1]{}
\fi

\begin{document}



%
%
%
%
%

\iffullversion
\title{Mind Your Credit: Assessing the Health of \\the Ripple Credit Network\thanks{This work appears at WWW 2018. The project's website is 
https://pedrorechez.github.io/Ripple-Credit-Study/}}
\author{Pedro Moreno-Sanchez, Navin Modi, Raghuvir Songhela, Aniket Kate, Sonia Fahmy\\
Purdue University\\
email: \textsf{\{pmorenos, modin, rsonghel, aniket, fahmy\}}@purdue.edu}
\date{Version: \today}
\else
\title{Mind Your Credit: Assessing the Health of \\the Ripple Credit Network}
\author{Pedro Moreno-Sanchez, Navin Modi, Raghuvir Songhela, Aniket Kate, Sonia Fahmy}
\affiliation{%
  \institution{Purdue University}
}
\email{{pmorenos, modin, rsonghel, aniket, fahmy}@purdue.edu}


\fi

\iffullversion
\maketitle

\begin{abstract}

The Ripple credit network has emerged as a payment backbone with key 
advantages for financial institutions and the remittance industry.
Its path-based IOweYou (IOU) settlements across different (crypto)currencies 
conceptually distinguishes the Ripple blockchain from cryptocurrencies (such as Bitcoin and altcoins), 
and makes it highly suitable to an orthogonal yet vast set of applications in the remittance world for cross-border transactions and beyond.

This work studies the structure and evolution of the Ripple network since its inception, and investigates its vulnerability to 
devilry attacks that affect the IOU credit of linnet users'  \accs.  
We find that
about $13$M USD are at risk in the current \nw due to inappropriate configuration of 
the rippling flag on \links,  
facilitating undesired redistribution of credit across those links. Although the \nw has grown around a few highly connected hub (gateway) \accs that 
constitute the core of the network and provide high liquidity to \users, such a \link distribution results in a user base of around $112,000$ \accs that can be financially isolated by 
as few as $10$ highly connected gateway \accs. 
Indeed, today about $4.9$M USD cannot be withdrawn by their owners from the \nw due to \payroutes, a gateway tagged as faulty by the Ripple community.
Finally, we observe that stale exchange offers pose a real problem, and exchanges (market makers) have not always been vigilant 
about periodically updating their exchange offers according to current real-world exchange rates.
For example, stale offers were used by $84$ Ripple \accs to gain more than 
$4.5$M USD from mid-July to mid-August 2017. 
Our findings should prompt the Ripple community to improve the health of the network by educating its users 
on increasing their connectivity, and by appropriately maintaining the credit limits, rippling flags,
and exchange offers on their IOU \links.
\end{abstract}

\else

\maketitle
\fi



\section{Introduction}
\label{sec:intro}
The \nw~\cite{Armknecht2015, linkingwallets, ripple-website, fuggermoney} conceptually differs from the plethora of flourishing cryptocurrencies 
because it simultaneously allows \txs across traditional fiat currencies, cryptocurrencies
as well as user-defined currencies over IOU credit paths.
Its inherent capability to perform cross-currency 
 \txs 
 in a matter of seconds for a small fee in a publicly 
verifiable manner paves the way for 
reducing costs of financial institutions and the remittance industry by billions of dollars~\cite{ripple-santander-save}.
Given that, 
early embracers of Ripple~\cite{fidor-ripple, us-banks-ripple} 
have been recently followed by a wave of financial institutions 
worldwide~\cite{rbc-ripple, ripple-banks-timeline, japan-ripple, bank-australia, santander-uk, ripple-institutions}, 
including 12 of the world's top 50 banks~\cite{fortune-top-banks}, 
remittance institutions~\cite{earthport, saldomx} and 
online exchange services for cryptocurrencies~\cite{gatehub-gateway, bitstamp-gateway}.

Among early academic efforts, Armknecht et al.~\cite{Armknecht2015} and 
Di Luzio et al.~\cite{icdcs2017ripple} 
present basic statistics of the \nw usage such as \tx volume, and consider the centralized nature of the Ripple blockchain consensus process respectively.
Moreno-Sanchez et al.~\cite{linkingwallets,pathshuffle} 
focus on deanonymization attacks and privacy enhancing solutions for \users. 
Nevertheless, the \nw is yet to get its due attention similar to Bitcoin~\cite{Meiklejohn2013,Bonneau15,Ron2013,Barber2012,DonetDonet2014} from 
the academic community.
This is critical because the Ripple 
IOweYou credit network
and path-based \txs over credit links clearly set it apart (structurally and functionally) 
from cryptocurrencies. 

 
Security of the credit in the \nw has not been studied thus far. Yet, it is crucial at this juncture to determine 
how users are handling their credit in the \nw and, more importantly, identify potential vulnerabilities, and 
determine countermeasures and best practices for future usage. 
By analyzing
the collected \nw data which includes 
$181,233$ \accs and $352,420$ 
\links, as well as $29,428,355$ 
\txs during the period Jan '13 -- Aug '17, 
we make the following key contributions.

This work presents 
the first extensive, longitudinal study of the \nw and its transactions throughout its \emph{complete} lifetime up to August 2017,  
shedding light on its evolution and analyzing its security.
We characterize the \nw graph (\cref{sec:structure}).
We show that the number of \accs and \links has grown
at a steady rate through 2016 with a sudden spike in 2017, in tune with wide adoption over the second quarter of 2017. The ratio between \accs and \links 
has however remained constant and hence the network density is decreasing. 
The network is slow-mixing, unclustered and disassortative. We identify \emph{gateway} nodes as the key players in the network today. 
Gateways are highly connected bootstrapping \accs trusted to set up links to new users. 
We show that \accs are dynamically grouped into geographically demarcated communities, where 
each community is defined by (on average) two gateway \accs. 
We find that the core of the \nw provides enough liquidity for transactions from other \accs.

We assess the security of the credit held by  \users in the \nw in three ways. 
First, we investigate the effect of 
undesired redistribution of credit, i.e., rippling (\cref{sec:rippling}). We show that 
more than $11,000$ \accs in the \nw are prone to rippling   
among their \links if they are used in a transaction as intermediate \accs. We observe that \links at risk are 
associated with more than $13$M USD. 

Second, we study the \resil to disruptive \accs in the \nw (\cref{sec:resilience}) and 
observe that although the core of the \nw, composed of around $65,000$ \accs,  
is resilient to disruptive \accs, there exists a large user base of more than $112,000$ \accs that 
is prone to disconnection by as few as $10$ highly connected gateway \accs, 
and their credit (currently about $42$M USD) is at risk of being no longer connected to the main component of the \nw and thus of being stagnated.
In fact,
we delve into the effect caused by a disruptive \acc by analyzing the case of \payroutes, 
a gateway tagged by the Ripple community as faulty~\cite{payroutes-forum1}. We observe that as of Aug '17, 
more than $600$ \accs still 
have credit issued by \payroutes for around $4.9$M USD that is stagnated (and cannot get transferred) as 
\payroutes does not provide the rippling option for those \accs. 

Finally, we study the effect that stale exchange offers have on the credit of market makers 
and \users (\cref{sec:stale}). 
In particular, we observe that 
during a period of ten days in 2013, 
market makers put at risk around $250,000$ USD due to stale offers, 
and $24$ \accs were able to gain more than $7,500$ USD 
by taking advantage of those offers.  
We find that this effect, caused by stale offers, not only continues 
in the current \nw but is amplified. In particular, during a period of 
one month in 2017, market makers 
put at risk at least $500,000$ USD due to stale offers, and cunning users 
gained more than $4.5$M USD.  


Our work motivates the Ripple community to enhance the health of the network by educating users on improving their connectivity and
setting the upper limits of their credit links well below the default value. Additionally, we encourage the market makers to frequently update their offers in the \nw, 
according to the corresponding exchange rates in the real world. 




\section{Background}
\label{sec:background}

The Ripple blockchain has emerged in the landscape of financial networks as an alternative 
settlement backbone for financial institutions and the remittance industry. Ripple adoption is fueled by potential savings of
more than $20$ billion dollars per 
year~\cite{ripple-santander-save}. At the time of writing, Ripple's market capitalization 
is third, only behind  Bitcoin and Ethereum. 

\paragraph{The Ripple network}
With its roots in IOweYou credit networks~\cite{DeFigueiredo2005,fuggermoney,linkingwallets,Armknecht2015},
the \nw essentially is a weighted, directed graph 
where nodes represent \accs and edges represent \links between \accs. 
The non-negative weight on an edge ($\uid_1$, $\uid_2$) 
 represents the amount that $\uid_1$ owes to $\uid_2$.
By default, the credit on a link is upper-bound by $\infty$, 
but the \acc owner ($\uid_2$ in our example) can customize it. 
Additionally, each \acc is associated with a non-negative amount of \xrp. 
\xrp is the native currency in Ripple, initially conceived perhaps for users to pay a small fee 
per \tx towards curbing denial of service attacks 
and unbounded \acc creation (or Sybil attacks). 

\Cref{fig:ripple-network} depicts an excerpt of the \nw. Here, the \link \emph{Bitstamp $\rightarrow$ Alice} 
denotes that Bitstamp owes Alice $1$ USD, and there is no upper-bound for such \link. The 
\link \emph{Gatehub $\rightarrow$ Edward} denotes that Gatehub owes Edward $5$ USD, and such credit can increase 
only up to $100$ USD. 

\iffullversion
\begin{figure}[tb]
\centering
\includegraphics[width=0.5\columnwidth]{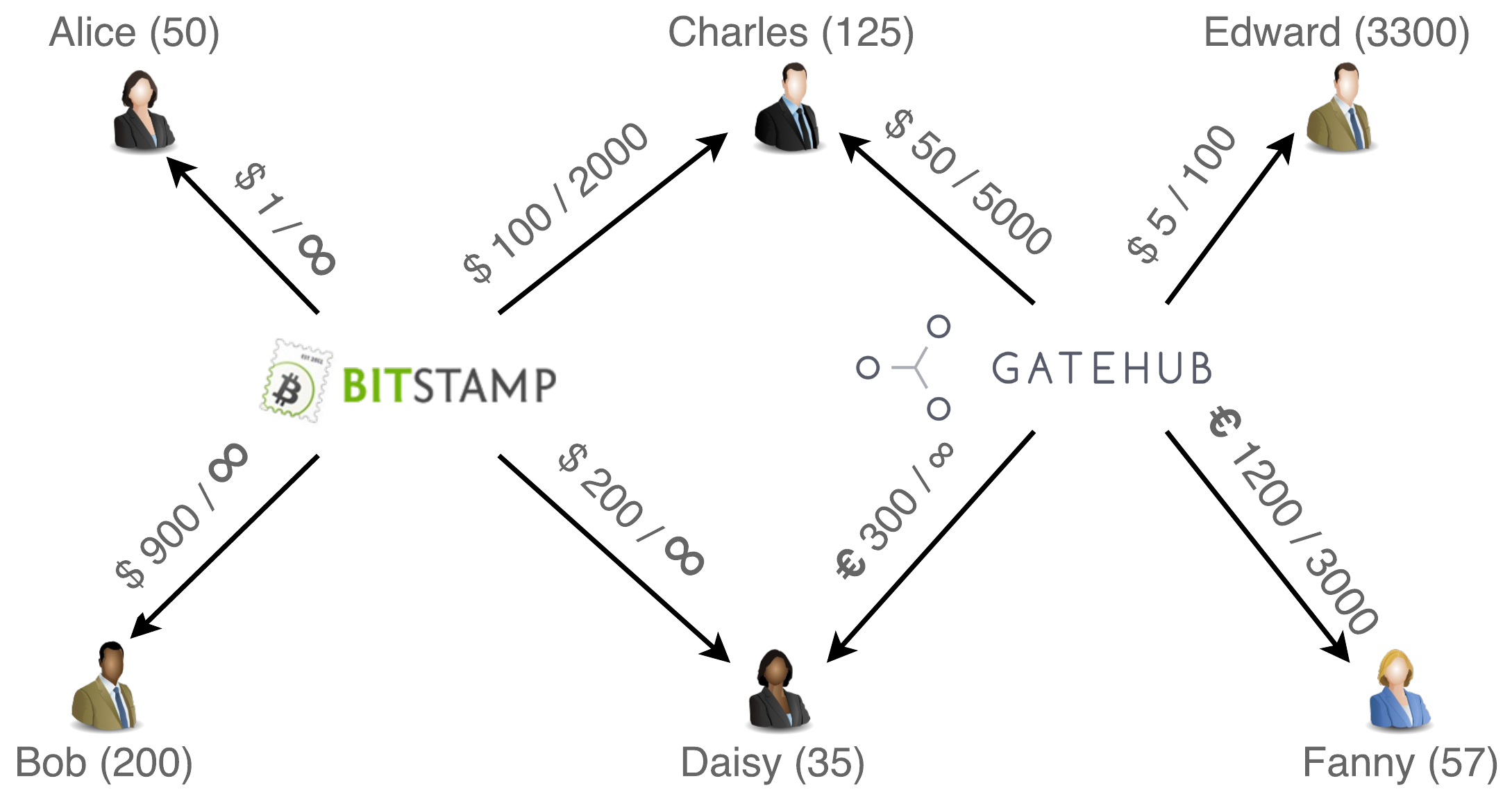}
\caption{Illustrative example of the \nw. Each \link is tagged with two values $a$ / $b$, where $a$ denotes 
the current credit and $b$ denotes the upper bound. The edge lower bound is always zero. Numbers within parentheses
denote the amount of \xrp owned by each user. 
\label{fig:ripple-network}}
\end{figure}
\else
\begin{figure}[tb]
\includegraphics[width=\columnwidth]{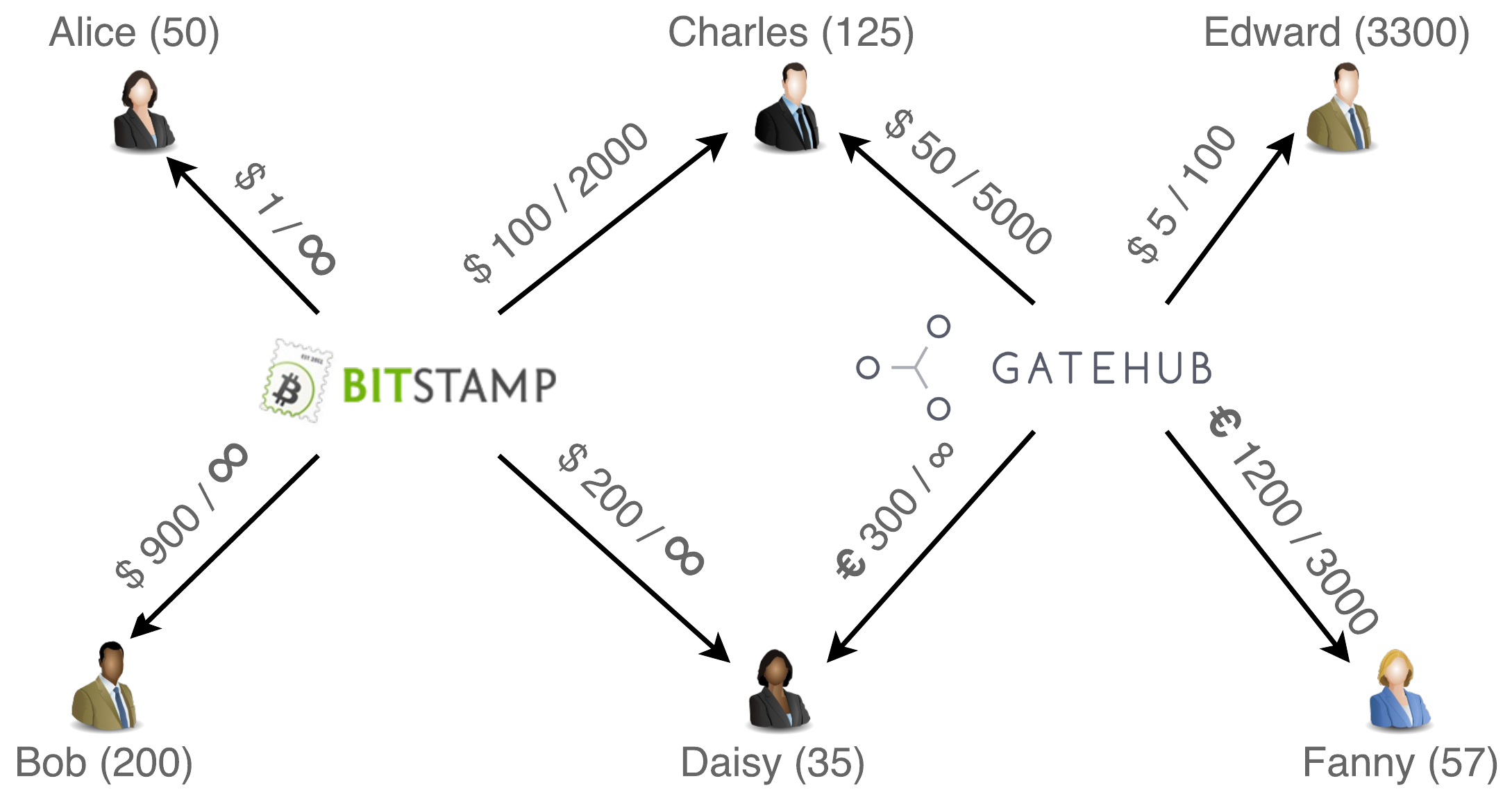}
\caption{Illustrative example of the \nw. Each \link is tagged with two values $a$ / $b$, where $a$ denotes 
the current credit and $b$ denotes the upper bound. The edge lower bound is always zero. Numbers within parentheses
denote the amount of \xrp owned by each user. 
\label{fig:ripple-network}}
\end{figure}
\fi

\paragraph{Ripple \accs} A \acc is governed by a pair of signing and verification keys from ECDSA 
or Schnorr signature scheme. An encoded 
version of the hash of the verification key identifies the \acc. Any operation associated with a \acc is only valid 
if it is signed by the corresponding signing key. Therefore, whoever holds the signing key for a \acc can do   
transactions with such \acc set as sender, create exchange offers or update its \links. 

\paragraph{Ripple transactions}
Ripple allows two types of \txs: direct \xrp payments and path-based settlement \txs. 
A direct \xrp payment exchanges \xrp  between two \accs, even 
if they are not connected via a network path. 
The payment amount is subtracted from the sender's \xrp balance and 
added to the receiver's \xrp balance.
Direct \xrp payments thereby resemble debit payments between \users rather than path-based credit settlements, 
which are the focus of this paper.
Therefore, we omit direct \xrp payments in our analysis and refer to~\cite{RipplestoryBitMEX} 
for more details. 

A path-based settlement \tx (or simply a \tx hereby) uses a path of \links between \sdr and \rcv to settle credit between them. 
In the example of~\cref{fig:ripple-network}, assume that Alice wants to pay Edward $1$ USD. 
At first, \links are considered undirected to find a path from the \sdr 
to the \rcv. The \tx can be routed using the path Alice $\leftarrow$ Bitstamp $\rightarrow$ Charles $\leftarrow$ GateHub $\rightarrow$ Edward.  
The \tx is carried out by updating the credit value on each \link depending on its direction as follows: 
\links in the direction from sender to receiver are increased by $1$ USD, while reverse \links are decreased by $1$ USD. 
In the running example,  Alice $\leftarrow$ Bitstamp 
and Charles $\leftarrow$ GateHub are decreased to $0$ and $49$ USD, respectively, whereas  Bitstamp $\rightarrow$ Charles 
and GateHub $\rightarrow$ Edward are increased to $101$ and $6$ USD, respectively. Several paths 
between \sdr and \rcv can be used in a single \tx~\cite{linkingwallets}.

\paragraph{Key players: Gateways and market makers}
A \emph{gateway} is a well-known business \acc established to bootstrap 
\links to new \accs in an authenticated manner. 
Gateways are the Ripple counterparts of user-facing banks and loan agencies in the physical world.
Their \accs maintain high connectivity. A newly created Ripple \acc
that does not initially trust any existing \acc can create a \link 
to a gateway and thereby interact with the rest of the network before forming
direct links to other \accs. Bitstamp and Gatehub are two examples gateways in the current \nw. 

A \emph{market maker} is a \acc that receives a certain currency on one of its \links and exchanges it for another currency on another \link, 
charging a small fee. Market makers enable \txs where senders and receivers hold different currencies. For instance, 
in~\cref{fig:ripple-network} assume that Bob wishes to pay $100$ EUR to Fanny by spending 
$120$ USD. Further, assume that Daisy has published an exchange offer USD/EUR at a rate $1.2$ USD $= 1$ EUR. 
In this manner, Daisy plays the role of market maker and facilitates the transaction from Bob to Fanny: Bob $\leftarrow$ Bitstamp is decreased by $120$ USD 
while Bitstamp $\rightarrow$ Daisy is increased by $120$ USD. Now, Daisy's offer is replenished, Daisy $\leftarrow$ Gatehub is decreased by $100$ EUR 
and finally, Gatehub $\rightarrow$ Edward is increased by $100$ EUR. 

\paragraph{Key operations: Rippling and exchange offers}
In the Ripple community, \emph{rippling} denotes the redistribution of credit on the links for each 
intermediate \acc as a consequence of a \tx~\cite{ripple-flag}. Rippling can only occur between 
two \links that belong to the same \acc and have credit in the same denomination. Nevertheless, 
several rippling operations can be concatenated to carry out a \tx with several intermediate \accs.

For instance, in \cref{fig:ripple-network} consider a \tx from Bob to Edward through Charles for a value of $40$ USD. 
Among other changes, this \tx decreases    
the balance of the link Bitstamp $\rightarrow$ Bob to $860$ USD and increases the balance 
of Charles $\leftarrow$ Bitstamp to $140$ USD, so that $40$ USD are shifted between the links of Bitstamp  
due to rippling. We expect that rippling is allowed by gateways; 
however, less active users may opt for avoiding balance shifts not initiated by them.  


An \emph{exchange offer} is created by a \acc to indicate its willingness 
to exchange one currency for another. Such \acc is then identified as \emph{market maker}. 
The typical exchange offers are of the 
type described above, where Daisy offers an exchange USD/EUR. However, 
the \nw also allows offers that involve \xrp. 
In fact, throughout the \nw lifetime, several market makers  
have included \xrp in their offers, later fulfilled by \accs as part of a path-based \tx. 

The combination of direct \xrp payment and path-based \tx is natively supported in 
the \nw and they are atomically executed as a whole. 
As they involve the reallocation of credit among \accs, we consider this 
type of \tx in this work and denote it as a \emph{path-based \tx involving \xrp}. We stress, however, that this is different 
from single direct \xrp payments, where only \xrp is involved.  
For instance, a path-based \tx involving \xrp can be used by a \acc 
to pay other \accs for performing \txs on its behalf.  Assume now that in~\cref{fig:ripple-network}, 
Fanny wants to pay Edward $1$ USD. However, she only has credit in EUR and Gatehub has not 
indicated any exchange offer of the form EUR/USD. Instead,  assume that Alice publishes an exchange 
offer of the type \xrp/USD. In such a situation, Fanny can pay the amount of \xrp corresponding to 
$1$ USD to Alice, who in turn transfers $1$ USD in the credit path connecting her to Edward. 

\section{Datasets}
\label{sec:data}

\paragraph{Data sources}
Our experiments are based on publicly accessible data  extracted through the API~\cite{ripple-api} provided by the Ripple network on 
their servers \emph{\{s1, s2, data\}.ripple.com}. We crawl the datasets describing the \nw topology (\accs and \links among them), transactions, gateways  
and market makers. We summarize these datasets in the rest of this section and refer the reader to~\cite{ripple-charts} for further statistics. 
The scripts for the data crawling and experiments are available at~\cite{ripple-myc}. 

\paragraph{Ripple network topology}
\label{sec:ripplenetwork}
We collected  all \accs and \links comprising the \nw at the end of each year from 
December 2013 
until December 2016, as well as at the end of August 2017. 
We model each snapshot as a directed graph with multiple edges between \accs (i.e., one edge per currency). 
For each snapshot, we only consider its largest connected component, and 
denote it by \g{year}. We observe that \g{17} consists of $181,233$ \accs 
and $352,420$ \links, which represent $98.6\%$ of \accs and 
$99.28\%$ of \links of the total \nw at that point of time.  Such percentages are similar for other 
snapshots. Therefore, we believe that \gall are representative of the \nw snapshots.


\paragraph{Ripple transactions}
We extract the \txs in the \nw in the period Jan '13 -- Aug '17, obtaining 
a total of  $29,428,355$ \txs. 
We prune this dataset according to the following criteria. 
First, we discard $1,530,107$ 
anomalous \txs (e.g., spam) considered outliers by previous studies~\cite{linkingwallets}. 
Second, we discard $16,180,972$ 
\txs carried out among \accs not included in \g{17}. The majority of 
these transactions are \xrp payments  
that do not require a credit path. 
Third, we discard $3,255,837$ direct \xrp payments among \accs in \g{17}. We  
only consider path-based transactions, even if they are extended with a 
\xrp payment.  
Our final set of \txs contains a total of $8,461,439$ 
\txs. We refer to this as \txdataset. 

\paragraph{Market makers}
We compiled the list of market makers present in \g{17} obtaining 
a set of $8,105$ \accs with at least one currency-exchange offer. 
We denote this dataset as \mm{17}. 

\paragraph{Gateways}
We crawled the list of gateways from the Ripple API, 
and added the gateways identified by the Ripple community throughout  
the \nw lifetime.
As a result, we obtained a list of $101$ gateways and $119$ \accs associated with them. 
We denote this dataset by \gwdataset. 


\paragraph{Ethical considerations}
Our \nw analysis solely uses publicly available data. Moreover, we do not deanonymize 
any user that owns a \acc in the \nw or include sensitive data about them. 
We only give the names of 
gateways that are well-known in the Ripple community, 
and publicly advertised on websites and forums.


\section{Graph Characteristics of the Ripple Network}
\label{sec:structure}

In this section, we dissect our datasets to investigate the structure and evolution of the \nw throughout its lifetime. 

\paragraph{\nw topology} 
\Cref{table:graph-overview} shows the \nw 
\accs and \links as well as the evolution of 
standard graph metrics for \gall. We make two observations. 
First, apart from the natural spike in the size of \g{14} due to the early stage 
of the system, \accs and \links have grown in 
\g{14-16} at a steady rate of 
$1.55 \pm 0.03$ and $1.52 \pm 0.07$ correspondingly, 
a trend showing that  \accs and \links grow at a similar rate  
and new \accs enter the \nw 
by connecting to a few existing \accs. However, these ratios 
have soared in the first eight months of \g{17}. 

\iffullversion
\begin{table}[b]\small
\centering
\caption{Graph metrics for the \nw topology for different snapshots. \label{table:graph-overview}}
\begin{tabular}{l | a c a c a}

 & {\bf \g{13}} & {\bf \g{14}} & {\bf \g{15}} & {\bf \g{16}} & {\bf \g{17}}\\ \hline
 \# \accs & $14657$ & $40051$ & $61173$ & $96953$ & $181233$ \\
 \# \links &  $26969$ & $82305$ & $119790$ & $190675$ & $352420$ \\
 \hline
 \hline
Avg degree      & $3.68$ & $4.11$ & $3.91$ & $3.93$ & $3.88$ \\ 
Clustering     & $0.08$ & $0.08$ & $0.08$ & $0.13$ & $0.07$     \\ 
Assortativity    & $-0.23$ & $-0.15$ & $-0.13$ & $-0.16$ & $-0.13$      \\
Density      & $12 \cdot 10^{-5} $ & $5.1 \cdot 10^{-5} $ & $3.2 \cdot 10^{-5} $ & $2.0 \cdot 10^{-5} $ & $1.0 \cdot 10^{-5} $      
\end{tabular}
\end{table}

\else

\begin{table}[b]\small
\centering
\caption{Graph metrics for the \nw topology for different snapshots. \label{table:graph-overview}}
\resizebox{\columnwidth}{!}{%
\begin{tabular}{l | a c a c a}

 & {\bf \g{13}} & {\bf \g{14}} & {\bf \g{15}} & {\bf \g{16}} & {\bf \g{17}}\\ \hline
 \# \accs & $14657$ & $40051$ & $61173$ & $96953$ & $181233$ \\
 \# \links &  $26969$ & $82305$ & $119790$ & $190675$ & $352420$ \\
 \hline
 \hline
Avg degree      & $3.68$ & $4.11$ & $3.91$ & $3.93$ & $3.88$ \\ 
Clustering     & $0.08$ & $0.08$ & $0.08$ & $0.13$ & $0.07$     \\ 
Assortativity    & $-0.23$ & $-0.15$ & $-0.13$ & $-0.16$ & $-0.13$      \\
Density      & $12 \cdot 10^{-5} $ & $5.1 \cdot 10^{-5} $ & $3.2 \cdot 10^{-5} $ & $2.0 \cdot 10^{-5} $ & $1.0 \cdot 10^{-5} $      
\end{tabular}
}
\end{table}
\fi

Second, we observe that most graph properties remain stable over the \nw 
lifetime except density,
which has continuously decreased after \g{14}. 
Since the ratio $\EE / \VV$ has been constant in 
the \nw, the density grows as $\frac{2}{|\VV| - 1}$, and therefore decreases as the number of \accs increases. 
This confirms the fact that the \nw is a sparse graph. 
We have validated these observations when considering 
the snapshot of the \nw every four months, considering thereby each economic quarter.

\paragraph{Ripple transactions}
We first separate \txdataset into two groups: (i) \Txs involving \xrp and (ii) 
\Txs not involving \xrp, obtaining  $1,751,394$  
\txs in the first group and  $6,710,045$ \txs in the second group. 
This shows that although \txs involving \xrp are supported in Ripple, they 
are not the norm. 

We make the following 
observations on Ripple \txs. First, we observe 
that there exist $2,001,650$ circular transactions where a \acc transfers credit to itself. 
Circular \txs can be used by a user, for instance,  to transfer credit from 
one gateway to another. Alternatively, as we discuss in \cref{sec:stale}, circular transactions 
are used by cunning users to gain credit from stale offers.
Second, we observe that there exist $2,136,387$ non-circular and cross-currency \txs that 
exemplify the use of the \nw for remittance. 
Third, we observe that there exist $2,608,891$ \txs that use at least one exchange 
offer available in the \nw. This demonstrates the importance of exchange offers.

Finally, we study the use of intermediate \accs in  payment paths. 
We count $1,285,024$ \txs that use $0$ intermediate \accs and represent mainly 
 deposit or withdrawal of credit with gateways. 
Moreover, we count $4,464,027$ \txs 
that use a single intermediate \acc and represent, among others, interactions of gateways with their 
users following the hot-cold \acc mechanism~\cite{linkingwallets}. 
Finally, we observe $2,712,388$ \txs that use two or more 
intermediate \accs and represent cross-currency \txs. They exemplify the use of 
rippling and exchange offers in the \nw.

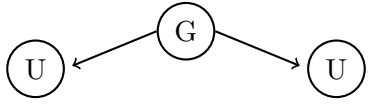
\begin{figure}[tb]
\centering
\begin{tikzpicture}[->,shorten >=3pt,auto, thick]
\node[draw, fill=white,circle,inner sep=2pt, text width=0.5cm, align=center, yshift = 0.5cm] (1) {G};
\node[draw,xshift=2cm,fill=white,circle,inner sep=2pt, text width=0.5cm, align=center] (2) {U};
\node[draw,xshift=-2cm, fill=white,circle,inner sep=2pt, text width=0.5cm, align=center] (3) {U};
\path (1.east) edge (2.west);
\path (1.west) edge (3.east);
\end{tikzpicture}
\caption{Most frequent motif in the \nw. U denotes user and G denotes 
gateway.\label{fig:motifs}}
\end{figure}

\paragraph{\nw structure}
Recent work~\cite{gleich-motifs,motif-wernicke} shows 
that high-order connectivity patterns or \emph{motifs} (i.e., a subgraph composed of three nodes connected via a certain pattern of two or 
three edges)
are important in understanding the structure of a graph. 
We follow this strategy to study the \nw structure. For that, we first  
classify \accs in \g{17} into gateways, market makers and \users, and 
color them accordingly. 
Using the \emph{FANMOD} tool~\cite{wernicke2006fanmod} on a colored version of \g{17} and parameters set 
to full enumeration, we find that 
the motif depicted in~\cref{fig:motifs} is the most frequently occurring with a frequency of $67.8\%$. 
This shows that  the \nw has gateways as key players, 
which is consistent with the low clustering coefficient and disassortativity properties in~\Cref{table:graph-overview}. 

\iffullversion
\begin{figure}[tb]
\centering
\includegraphics[width=0.5\columnwidth]{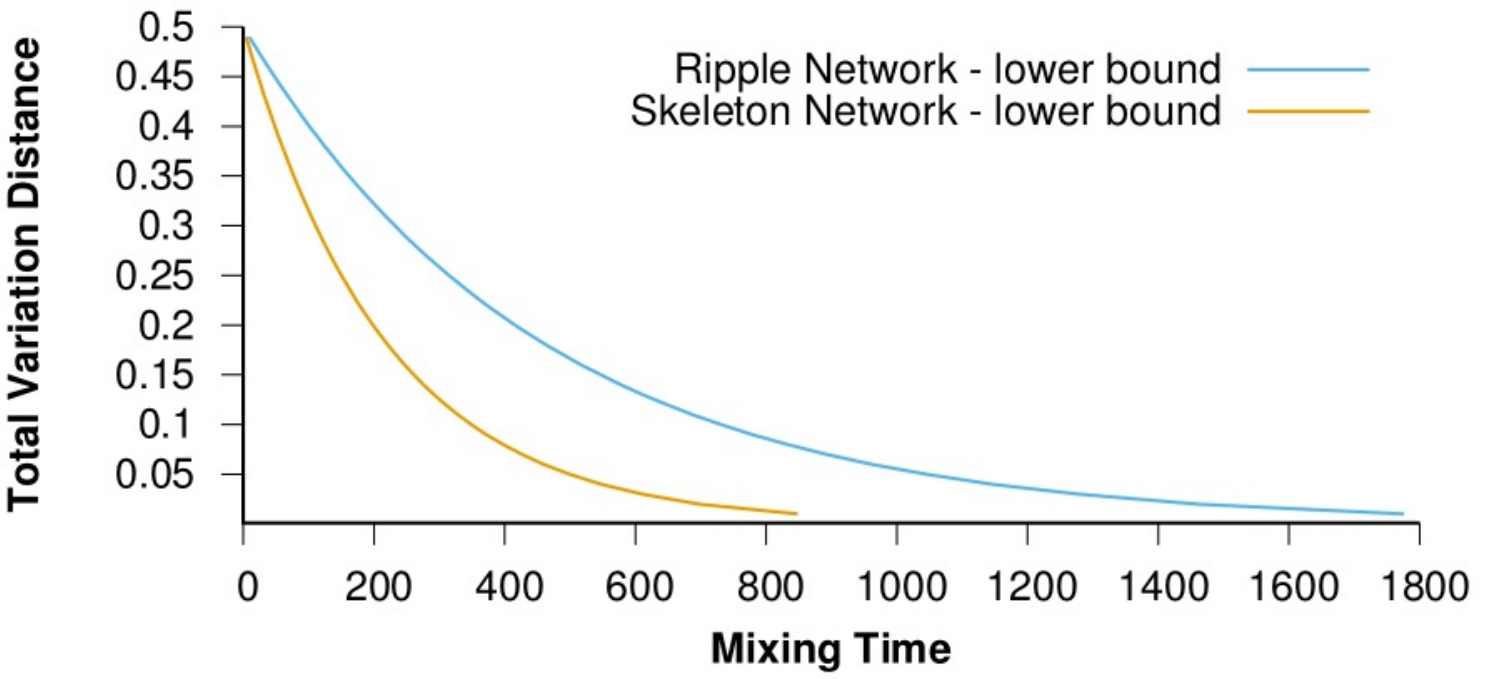}
\caption{Lower bound on the mixing time for \g{2017} and \s{2017}. \label{fig:mixing-time}}
\end{figure}
\else
\begin{figure}[b]
\includegraphics[width=\columnwidth]{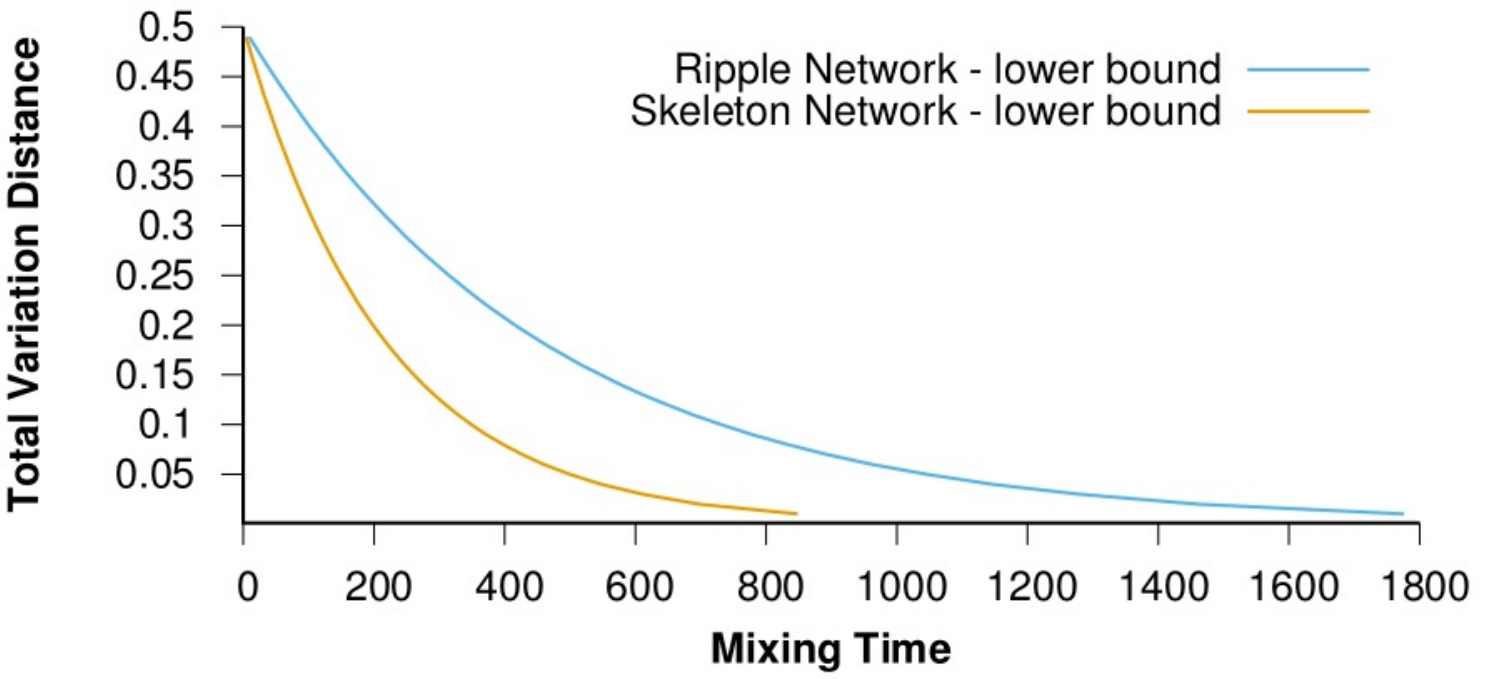}
\caption{Lower bound on the mixing time for \g{2017} and \s{2017}. \label{fig:mixing-time}}
\end{figure}
\fi
\paragraph{Mixing time}
\label{sec:mixing}
As described by Mohaisen et al.~\cite{Mohaisen-mixing}, the mixing time in a graph represents how quick a random walk on the graph reaches the stationary
distribution. In terms of a payment network as the \nw, mixing time intuitively determines how many intermediate 
\accs are required to reach a receiver \acc from any given sender \acc.   

We compute a lower bound on the mixing time of the \nw using the second largest Eigenvalue of the transition matrix for the graph as
described in~\cite{Mohaisen-mixing} (\cref{fig:mixing-time}).
We make two observations. First, the lower bound on the mixing time 
for an $\epsilon = 0.10 $ is $730$. 
This slow-mixing property is similar to the one observed for social networks~\cite{Mohaisen-mixing}, which is not surprising given the small clustering coefficient observed.  
The fact that the \nw is slow mixing increases the need for intermediate \accs to perform a transaction between any two \accs. 
Second, the mixing time decreases if we consider the core of the \nw (\s{17}), a phenomenon also 
observed for social networks~\cite{Mohaisen-mixing}. Here, \s{17} is obtained by iteratively removing \accs with 
a single neighbor from \g{17}. This shows that the core of the \nw has higher connectivity than the periphery.



\paragraph{Communities in the \nw} 
We next consider how \accs group into communities and how those 
communities have evolved over time. 
We extract the communities using 
the Louvain community detection algorithm~\cite{Louvain2008} as 
implemented in the Gephi software~\cite{gephi} on input \g{17}. 
The Louvain algorithm is parametrized by a \emph{resolution} to 
determine the granularity in the search for communities. We set 
this parameter to $0.45$ since we observe that lower values of 
the resolution result in 
smaller communities that trivially form around a single gateway, whereas higher 
values result in larger communities containing several 
gateways that may be geographically located far apart. 
For the chosen parameter, we have extracted  $77$ communities of sizes ranging 
from $3$ to $23869$ \accs. 

We then derive the geographical location for each community to shed light on the 
community structure of the \nw. We opt for geographical location since most gateways 
require users to provide identity and address verification documents 
before they populate links to them and may restrict users by geographical location. Towards this goal,  
we first map each gateway included in \gwdataset to its geographical location based on 
the information included in their corresponding websites; and second, 
we map a community to the location of the gateway(s) contained in the community, 
discarding $61$ communities that are not associated with a known gateway. The  
discarded communities are the smallest communities we found, with sizes ranging from 
$3$ to $999$ \accs.


\iffullversion
\begin{figure}[tb]
\centering
\includegraphics[width=0.7\columnwidth]{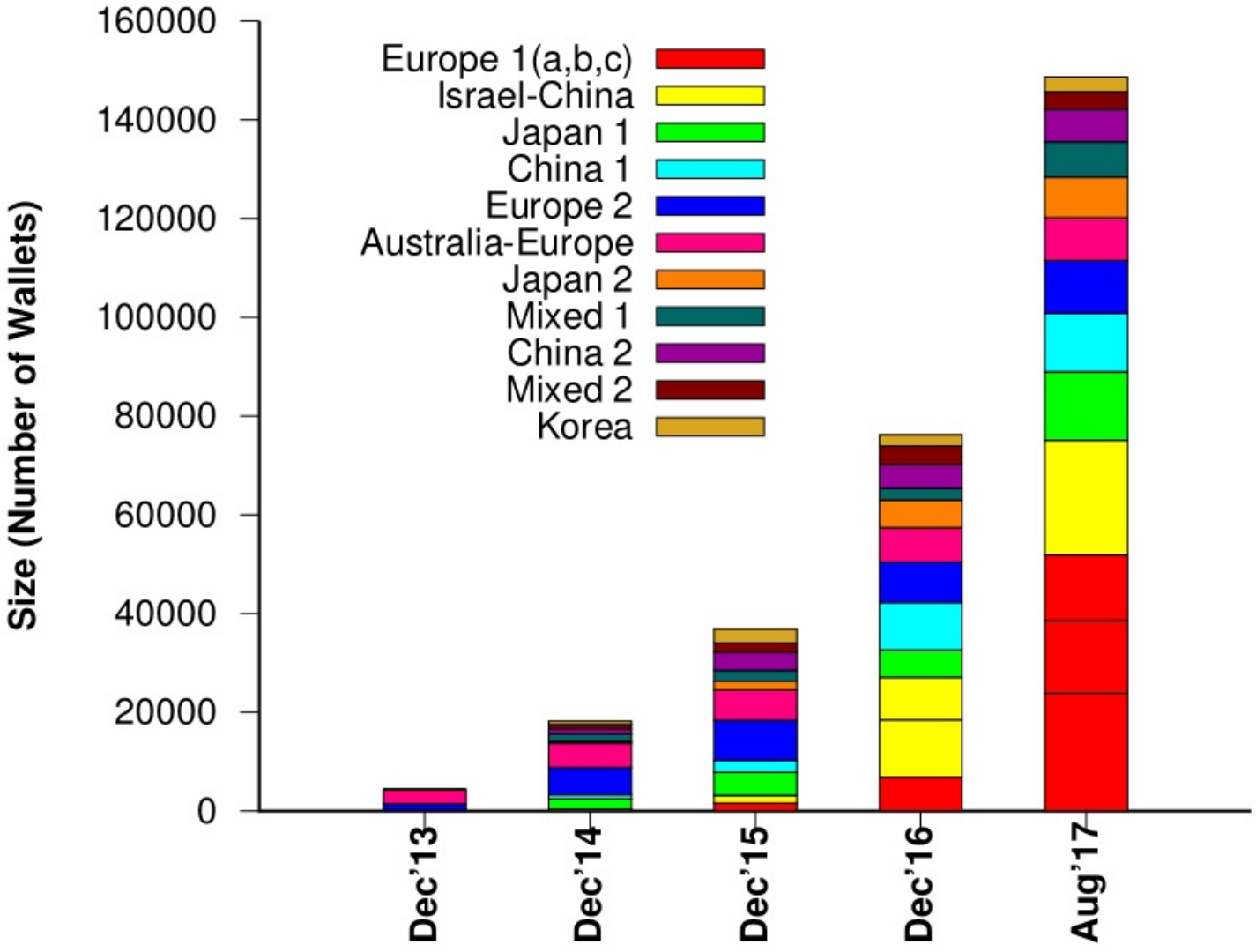}
\caption{Distribution of communities over time. Each stack shows the size of the community 
in that snapshot. Communities located in the same region are additionally labeled with a number. 
Finally, split or merge of communities are represented by split stacks of the same color. \label{fig:comm-tl}}
\end{figure}
\else
\begin{figure}[tb]
\includegraphics[width=\columnwidth]{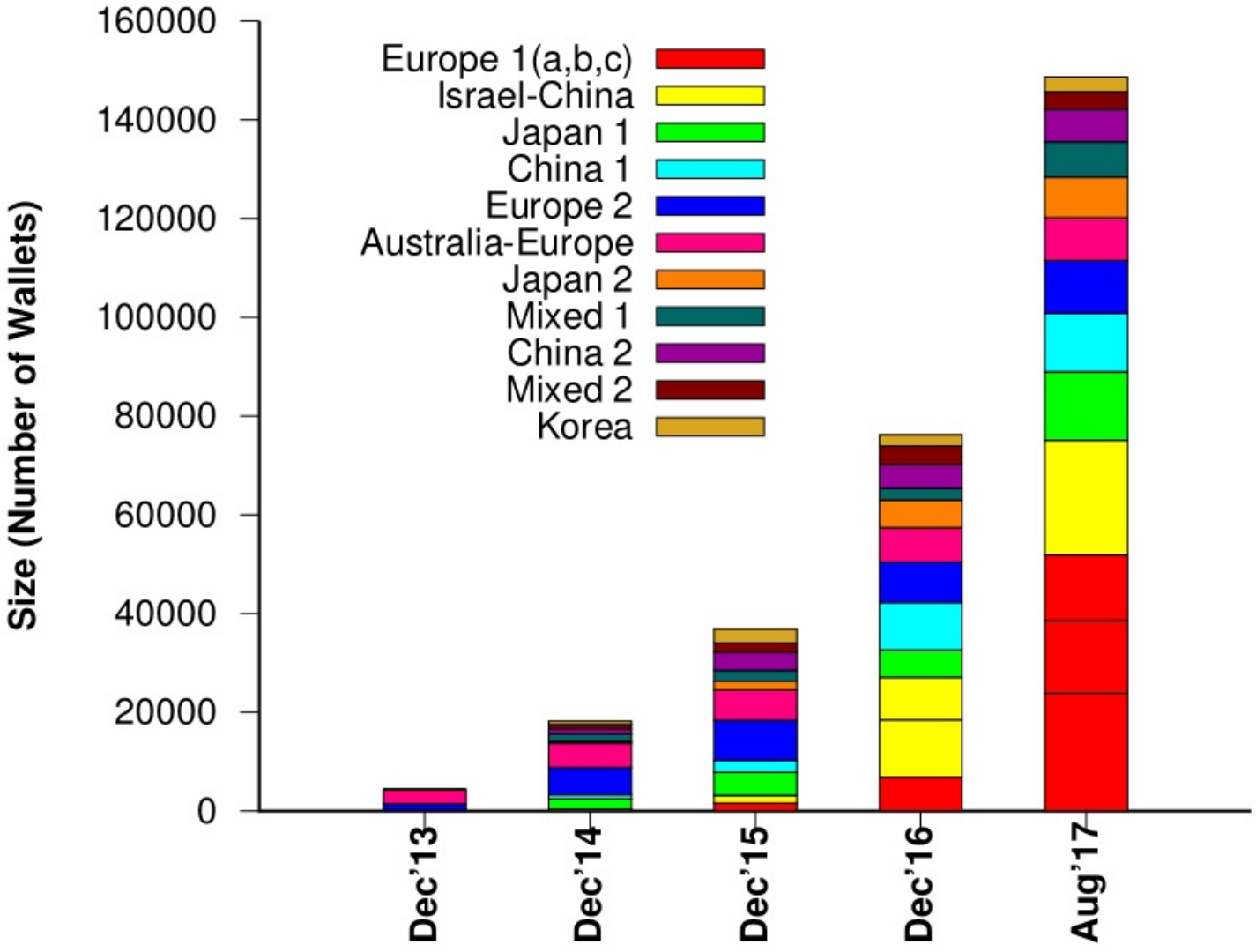}
\caption{Distribution of communities over time. Each stack shows the size of the community 
in that snapshot. Communities located in the same region are additionally labeled with a number. 
Finally, split or merge of communities are represented by split stacks of the same color. \label{fig:comm-tl}}
\end{figure}
\fi

\cref{fig:comm-tl} depicts the results of this experiment. 
Mixed-1 refers to Japan, HongKong, Turkey, NewZealand, and Mixed-2 refers to Sanghai, Canada, Indonesia, Singapore, Latin America. 
As of {Aug '17}, the largest community centers around Gatehub (Europe 1(a) in~\cref{fig:comm-tl}), a key gateway in Europe, followed by 
a community represented by \payroutes and RippleChina. 

To understand the evolution of the communities, we  repeat the experiment with 
\g{13-16}.  We  observe that 
communities are dynamic. In fact, the community tagged as Europe-1 in Dec '16 has been split into 
three communities Europe1(a,b,c) as of Aug '17, built around three emerging gateways in Europe. Conversely, 
communities separately built around \payroutes and RippleChina in Dec '16 have merged together in Aug '17. We believe 
that this phenomenon corresponds to the growing activity between \accs of these two gateways.

In summary, Ripple user communities form by connecting to gateways in the same geographical region. This is a result of the identity verification process enforced by many gateways. Despite the pseudonymous nature of Ripple \acc identities, 
this geography of communities can simplify identification tasks for regulation and law-enforcement authorities. 
However, the identification process before a new \link is created and funded reduces the number of \links in the \nw. This results 
in a slow-mixing, unclustered, disassortative network. The slow-mixing property is similar to other networks where link creation requires physical interaction~\cite{dellAmico2009}.


\paragraph{Ripple liquidity}
We say that 
a pair of \accs (\sndr, \rcv) has \emph{liquidity} if the amount of \credit that can be 
transferred between them is only bounded by the \credit available on either the \sdr or 
\rcv \links.  We now study whether credit in the \nw effectively facilitates  
transactions among Ripple \accs. 
First, we prune from \g{17} the \links associated with 
a currency other than \set{USD, CNY, BTC, JPY, EUR},
extract the largest connected component of the pruned graph, 
and convert the balance on the remaining \links to USD using publicly available exchange rates. 
We select these five currencies since they are the most common, comprising more than  
$65\%$ of the original \links. 
We denote this processed subgraph as \prung{17}. 

\iffullversion
\begin{figure}[tb]
\centering
\includegraphics[width=0.5\columnwidth]{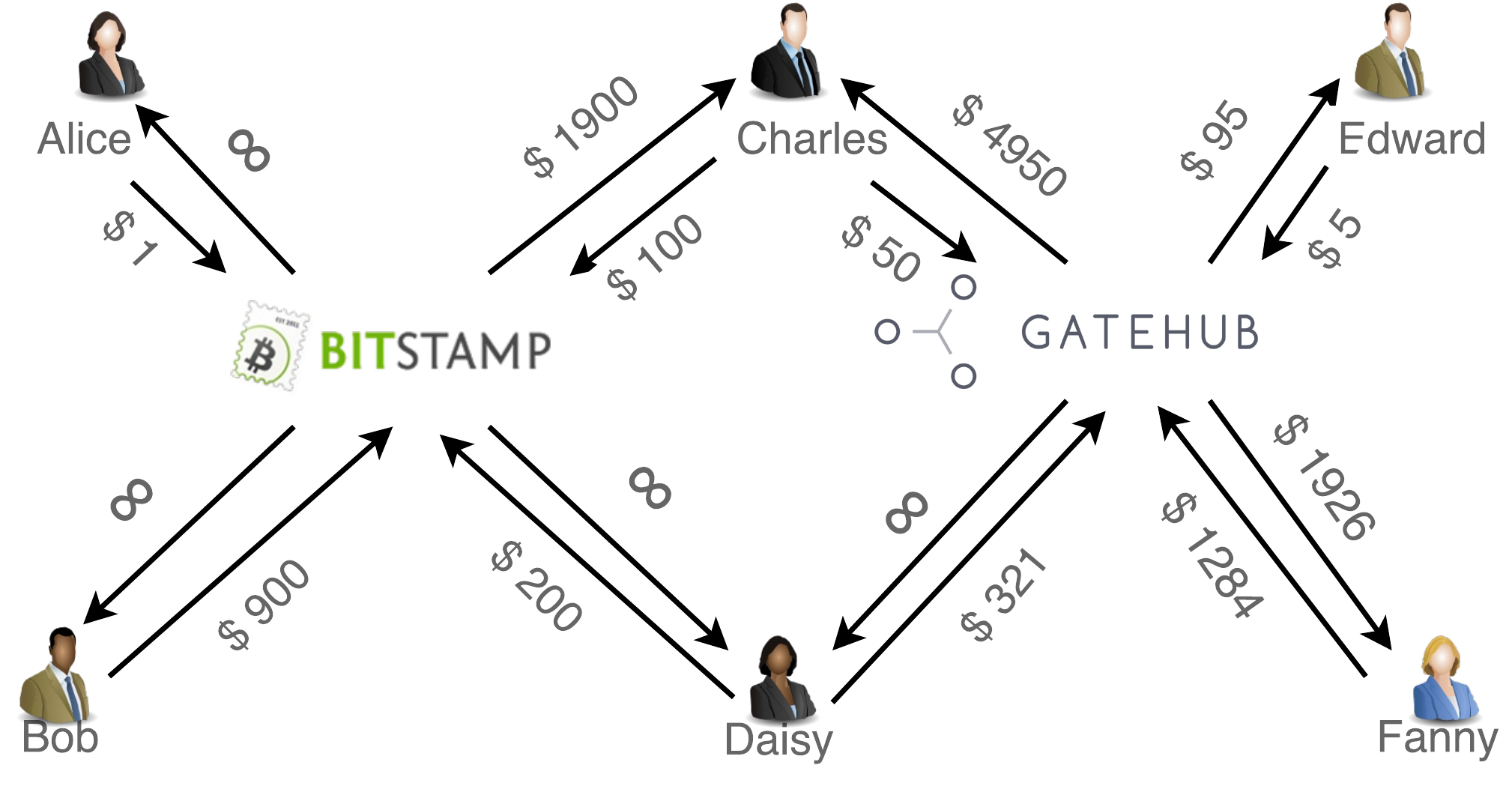}
\caption{Example graph for liquidity experiment. Each edge weight shows the credit that can be transferred from the source of the edge to its destination.
\label{fig:liquidity-flow}}
\end{figure}
\else
\begin{figure}[b]
\includegraphics[width=\columnwidth]{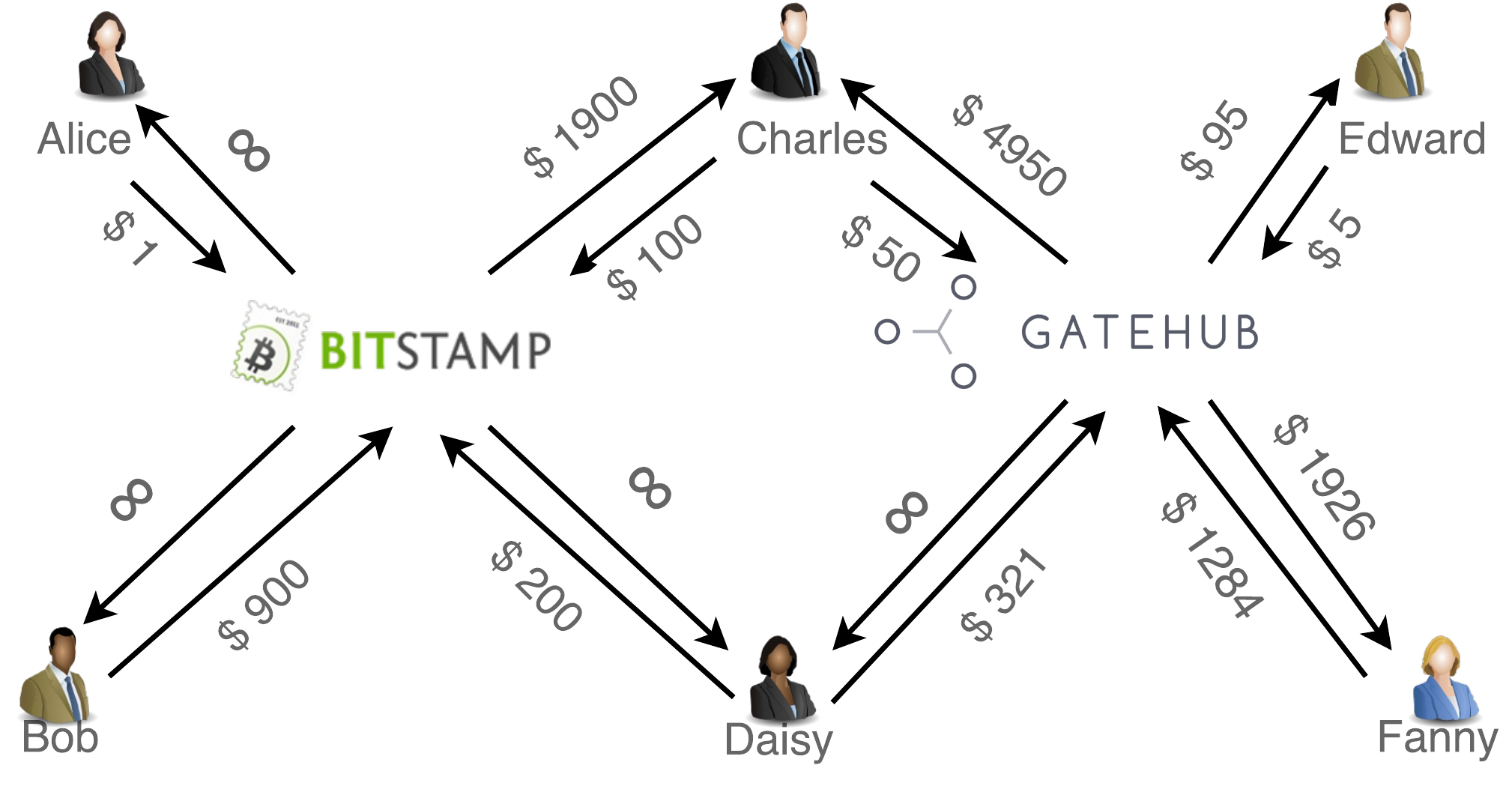}
\caption{Example graph for liquidity experiment. Each edge weight shows the credit that can be transferred from the source of the edge to its destination.
\label{fig:liquidity-flow}}
\end{figure}
\fi

Second, we transform \prung{17} to denote how much \credit can be transferred between \accs instead of 
how much \credit one \acc owes to its counterpart, as described by Dandekar et al.~\cite{Dandekar-liquidity}. 
For example, the \link Gatehub $\rightarrow$ Edward with balance \$5 and limit \$100 
in \cref{fig:ripple-network}, results in two \links: Gatehub $\rightarrow$ Edward with value \$95 and 
Gatehub $\leftarrow$ Edward with value \$5. Following this approach,  \cref{fig:ripple-network} can be transformed into 
\cref{fig:liquidity-flow}. 
We denote this transformed graph by \liqg{17}. 

Finally, we check the liquidity on \liqg{17} by randomly picking a representative 
sample of the \nw consisting of $10,000$ pairs of \accs 
avoiding repetitions, and for each pair
($\wid_1$, $\wid_2$), we calculate the max-flow from $\wid_1$ to $\wid_2$.
%
%
We observe that $92.55\%$ of the pairs of \accs have liquidity. In other words, 
the max-flow value between \accs is determined by the \credit value available on either $\wid_1$'s \links or 
$\wid_2$'s \links. 


In conclusion, the core of the \nw provides high liquidity and the bottleneck for 
\txs are the \links from the \users. In terms of liquidity, the \nw is 
similar to the current banking system, where the major
banks hold more \credit than their customers.

\section{Rippling and users: The effect of unexpected balance shifts}
\label{sec:rippling}

Although rippling maintains the net balance of intermediate \accs, its use is not 
innocuous for intermediate \accs.
%
The main issue is that the actual market value and stability of the \credit depends on the issuer of such \credit. 
In our illustrative example of \nw in~\cref{fig:ripple-network}, Charles may trust the credit from Gatehub more than Bitstamp.
Therefore, a \tx involving rippling among the two corresponding \links can induce a redistribution 
of \credit from a more valuable to a less valuable issuer without 
the specific consent of the involved \acc's owner. 
We expect gateways to allow rippling;  
however, less active users may wish to avoid balance shifts not initiated by them.

As a countermeasure, each \link is associated with a flag \rflag. When \rflag is set, 
the corresponding \link cannot be part of a rippling operation. This flag was 
first added in December 2013, and was updated in March 2015 to have a default state 
of ``set'' (i.e., no rippling allowed by default),  so \users could selectively opt-out and allow rippling. Additionally, a \acc has 
a new flag called \dflag that, if set, enables 
rippling among all the \acc's \links. Gateway \accs, for instance, follow this 
pattern~\cite{ripple-file}. 

\paragraph{Goal} In this experiment, we aim to identify \accs other than gateways 
that allow rippling, and to extract how much credit they put at risk doing so. 

\paragraph{Methodology} First, the \links not including \rflag flag are 
tagged as $\rflag = \text{false}$. Second, for each \acc that has the \dflag flag set, we set $\rflag = \text{false}$ (i.e., rippling is allowed) on all its \links. 
Third, we use the \rflag flag for the remainder of the links as specified in the \g{17} dataset. 
Now, we say that a \acc is prone to rippling if it has at least two \links 
with $\rflag = \text{false}$ (i.e., they allow rippling) and they hold credit in the same currency. 

\paragraph{Results} We find that more than $11,000$ 
\accs are prone to rippling and are not associated with well-known 
gateways. Moreover, more than  
$13$M USD are prone to rippling, counting 
only the \links that \accs prone to rippling have directly with gateways, as they are associated with real-world deposits. 
This gives a lower bound on the amount of credit at risk, and the actual value could be higher, if we count
credit at risk with \accs other than the gateways. 
This result demonstrates that unexpected balance shifts in the \nw can still affect a significant number of \accs, and more importantly, their credit.

We also observe that 
many \accs prone to rippling maintain \links with a low balance (even zero), 
but with upper limit set to a value larger than zero. 
The gap between balance and upper credit limit on these \links can be used to shift the balances of \accs, thus increasing the risk. 

\paragraph{Countermeasures} The users have the possibility of disabling the 
rippling functionality on their \links completely. Therefore, less active users may opt for disabling rippling among 
their \links to avoid balance shifts not initiated by them. Moreover,  more active users can also opt for 
dynamically adjust the amount of credit prone to rippling and add a rippling fee to it. Finally, users with 
\links holding zero balance should 
reduce their upper limit to effectively void them. 


\section{Rippling and gateways: The effect of faulty gateways}
\label{sec:resilience}

The gateway wallets are highly connected wallets included in the core of the Ripple network 
and significantly contribute to the liquidity of the network.
A faulty gateway can disable rippling on most \links of its 
\acc, ensuring that \txs routed through it are no longer possible and effectively 
freezing the balance held at \links of its \acc~\cite{ripple-bitstamp-freeze, RipplestoryBitMEX}.
This would not only severely affect the liquidity of the network, 
but also lead to monetary losses to the neighboring \accs, as they no longer can use 
the credit issued by the compromised \acc. 

\paragraph{Goal}
We aim to study the effect of 
faulty gateway \accs (e.g., as a result of adversarial \acc compromise) and the \resil of the \nw to them. 





\iffullversion
\begin{figure}[tb]
\centering
\includegraphics[width=0.5\columnwidth]{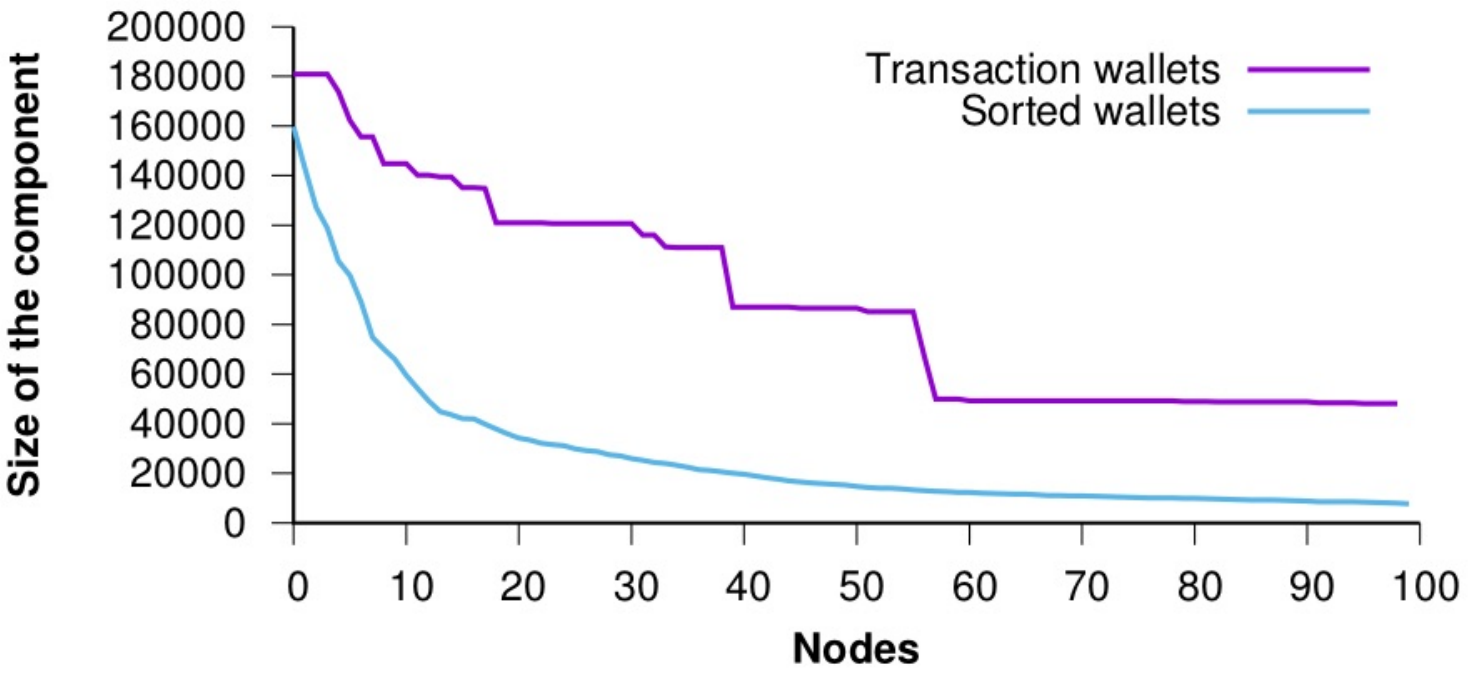}
\caption{Size of the largest connected component after removing \accs sorted by 
number of \links (blue) and number of appearances in \txs (purple).\label{fig:NetworkRobustness}}
\end{figure}
\else
\begin{figure}[b]
\centering
\includegraphics[width=\columnwidth]{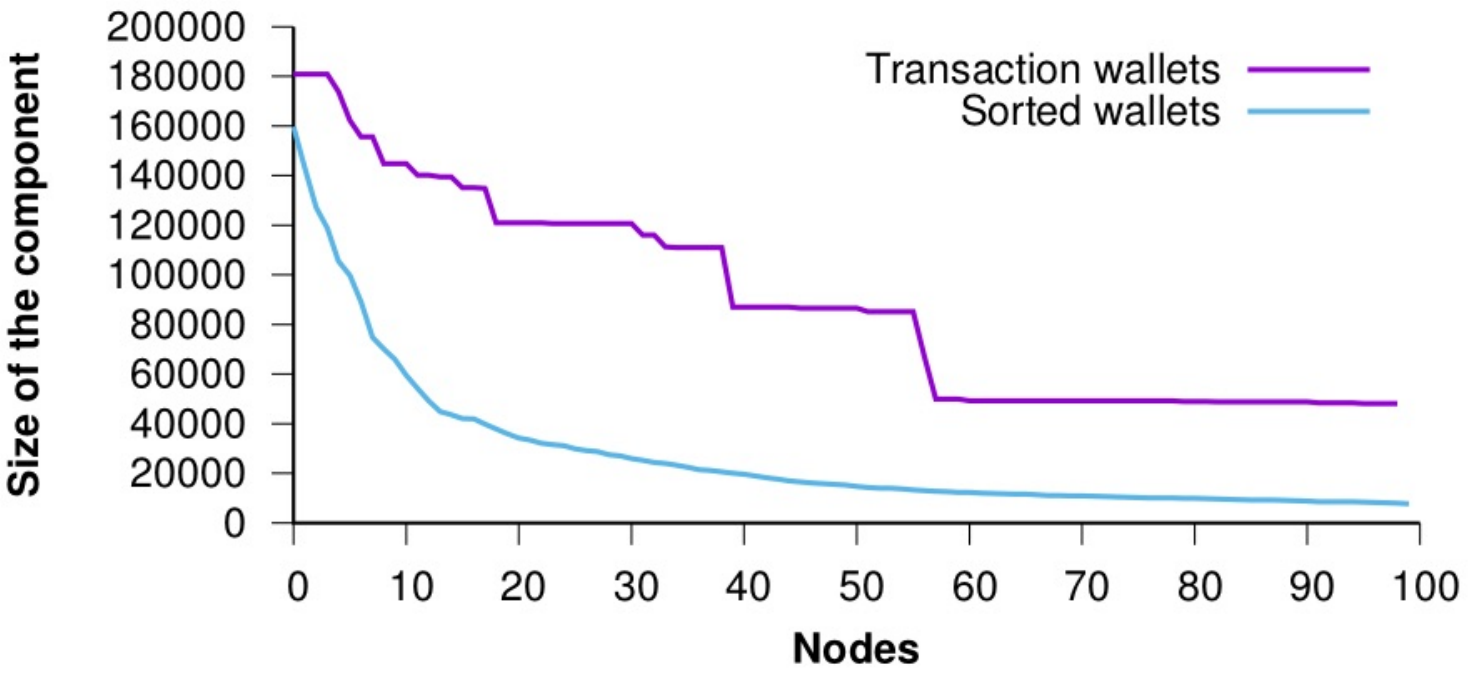}
\caption{Size of the largest connected component after removing \accs sorted by 
number of \links (blue) and number of appearances in \txs (purple).\label{fig:NetworkRobustness}}
\end{figure}
\fi

\paragraph{Methodology} We select $100$ candidate faulty \accs from \g{17} according to two different criteria: 
(i) \Accs with highest degree (\snodes) and (ii) \Accs involved in most of the \txs 
(\tnodes). 
%
We assess the most disruptive set of \accs by removing them from 
\g{17} and observing how the network connectivity is affected. 
%
\Cref{fig:NetworkRobustness} depicts the size of the largest connected component 
after removing the \accs in \snodes and \tnodes.  Intuitively, the 
smaller the component, the fewer the possible \txs, since only \accs 
in the same component can transact with each other. 
From this experiment, we conclude that \accs included in \snodes have a more profound impact 
on the connectivity of the \nw (and therefore on the \txs) than \accs included in \tnodes. Therefore, we use  \snodes in the rest of this section. 

We define the \emph{resilience factor} (\resilfactor) as the ratio 
between the component size in the most disruptive splitting of the network after removing a \acc (i.e., 
splitting the network in two components of equal size) and the size of the actual largest 
component after removing a \acc. 
%
%
%
%
Therefore, the \resilfactor can take values in the range $[0.5, 1]$. Values close to 1 indicate that the network has a low resilience, 
as the removal of a \acc resulted in a component with (close to) half of the \accs of the network. 
Conversely, values close to 0.5 indicate that the network has a high resilience, as the 
largest component after removing a \acc is (close to) the entire graph. 

\paragraph{Results} We observe that the \resilfactor of the \nw is maintained in the range $(0.5, 0.6)$ 
after the removal of each \acc in \snodes, demonstrating that the 
core of the \nw has high resilience. We conclude that we 
can divide the \nw into: (1) A small network core of around 
$65,000$ \accs ($36\%$ of the total) 
that includes the key \accs with high connectivity. This core is 
highly resilient to the removal of highly connected \accs, and (2) A large set of around 
$112,000$ \accs that 
can be easily disconnected from the network after removal of key \accs.
Yet, these highly vulnerable \accs 
have more than 
$42$M USD of credit with the gateways, 
which is at risk.

\paragraph{Countermeasures} This result shows that the \nw still has a few \accs that are ``too big to fail.'' 
As a countermeasure, it is necessary for many users to increase their connectivity and split their credit among 
different \links to avoid losses due to the failure of
a handful of \accs.

\subsection{A case study: The \payroutes gateway}

While studying the \nw communities (see~\cref{sec:structure}), we observed that the size of the
community created around the \payroutes gateway suddenly increased in Dec '16. 
Surprisingly, users in the Ripple community had reported the unresponsiveness of the 
company running the gateway when contacted regarding the credit issued by it~\cite{payroutes-forum1}.
We also emailed them, but got no answer at the time of this writing. In this state of affairs, we study \payroutes 
as an example of a faulty gateway. 

\paragraph{Goals} We consider two questions. First, we 
aim to find the amount of credit in the \nw that can only be withdrawn with 
the cooperation of \payroutes and, given the unresponsiveness of the gateway, 
this credit is ``stuck'' in the \nw. 
Second, we study why \accs with stuck credit obtained it in the first place, 
even though \payroutes was already reported 
as faulty. 
We describe our methodology and results for each goal separately in the following two sections. 
 
\subsubsection{Credit with \payroutes} 
\label{sec:acc-stuck-credit}

We are interested in \links of the form 
\payroutes $\rightarrow \wid_i$  where \payroutes has 
disabled rippling. This implies that the credit on these links can only be used in a withdrawal operation 
jointly with \payroutes: $\wid_i$ sets the credit on the link to $0$ to obtain the corresponding 
amount in the real world from \payroutes. However, as \payroutes is a faulty gateway, this operation is no longer available 
and the credit is stuck. 
Given that, we first address the question: \emph{how much credit is stuck on \links with 
\payroutes?} 

\paragraph{Methodology} From \g{17}, we first 
pick the \links with \payroutes as counterparty and positive balance, and derive the status of their rippling flag (as described in~\cref{sec:rippling}).  
Then, we classify the neighbor \accs of \payroutes into two groups 
as follows. First, we identify those \accs that have a \link with \payroutes for which rippling is not allowed, i.e., 
\rflag is set to true. We denote this set of \accs by \emph{\accs-no-rippling}. 
Second, we consider the set of \accs that are not in \emph{\accs-no-rippling} but yet cannot perform a transaction 
for an amount equal to the balance on their \link with \payroutes. We denote this second set as 
\emph{\accs-rippling-no-tx}. 
As the \accs in either \emph{\accs-no-rippling} or \emph{\accs-rippling-no-tx} cannot transfer the 
(entire) credit they have 
on a \link with \payroutes to another \acc in the \nw, the only way for them to get their credit back is to contact 
\payroutes in the real world and withdraw the corresponding funds. However, as \payroutes is unresponsive,  such credit is ``stuck.'' 

\paragraph{Results} We observe that, out of the $2,958$ \accs that have at least 
one \link with \payroutes,  there exist $621$ \accs in either \emph{\accs-no-rippling} or \emph{\accs-rippling-no-tx}, 
and therefore with stuck credit. We observe that the stuck credit on these \links is  
around $4.9$M USD. 

\paragraph{Discussion} The \payroutes case is not typical in the \nw.  
There have been other gateways that have ceased operation during the 
\nw lifetime, but have not caused such an effect. We consider \dr as an example of such a gateway. 
The difference from \payroutes is that before shutting down, \dr publicly announced it and mandated its clients
to proceed to withdraw the credit available in their \links with \dr.

We conduct the same above experiment for \dr, and
observe that, although $665$ \accs have credit stuck with \dr, such credit 
accounts for around 
$1,000$ USD only. This is how much \dr currently owes to the rest of \accs. 
This demonstrates that \accs followed the 
announcement of the gateway and successfully managed to withdraw most of their 
credit before the gateway ceased operation.

\subsubsection{Obtaining credit from \payroutes}

In this section, we focus on answering the question: \emph{How did \accs 
with stuck credit obtain such credit in the first place?}

\paragraph{Methodology} 
We first investigate how new \links were created with \payroutes over the 
lifetime of the \nw. 
We observe a spike of  
$2,527$ \links created in Oct '16 from a total of $1,805$ \accs. Out of these, $186$ \links 
were created by $133$ \accs and have  
balance stuck in \payroutes. This implies that $21\%$ of the \accs with stuck balance 
created \links with \payroutes during that month. We denote these by \stuckacc{Oct}{16}. 


Given this unusual behavior, we study how those $133$ \accs obtained credit. We identify two possibilities: 
(i) A path-based transaction from another \acc in the \nw; 
(ii) A circular transaction (i.e., sender and receiver of the transaction are the same \acc), where 
a \acc pays a certain amount of \xrp (or any currency issued by a gateway other than \payroutes)  
in exchange for credit issued by \payroutes on a \link with it. 

%

\paragraph{Results} We observe that \accs in \stuckacc{Oct}{16} do not receive significant credit from 
other \accs in the \nw during October 2016. In particular, we find only three transactions with credit values 
of $10$ USD, $100$ ILS and $5$ ILS.  Instead, \accs in \stuckacc{Oct}{16} get their credit through 
circular transactions. We find that $51$ \accs perform a total of $286$ circular transactions, where 
these \accs received around 
$12,000$ USD in exchange for 
approximately 
$300$ CNY and $12,000$ \xrp. 


In essence, \accs in \stuckacc{Oct}{16} invested mostly \xrp to obtain USD from \payroutes. 
We find that the exchange rate \xrp/USD 
in the \nw was considerably ``better'' than in the real world at that time: In the \nw at that time, a \acc could 
get $0.73$ USD for $1$ XRP on average, with a minimum of $0.14$ and a maximum of $2.87$ USD 
using stale offers available in the network. However, in the real world, one could get less than $0.01$ USD
for $1$ \xrp at the average exchange rate at that time and up to $0.28$ USD for $1$ \xrp, 
even considering the best exchange rate over the entire \nw lifetime.

\iffullversion
\begin{table}[tb]
\centering
\caption{Summary of the exchange offers between \xrp and USD created in the \nw during 
 October 2016.\label{table:offers-oct16}}
\begin{tabular}{c c | c c | c}
Pay Val & Pay Cur & Get Val & Get Curr & Ratio\\
\hline
$1062738.51$ & XRP & $17009.50$ & USD & $62.48$ to $1$\\
$59678.62$ & USD & $33194.62$ & XRP & $1.78$ to $1$\\
\hline
\end{tabular}
\end{table}
\else
\begin{table}[tb]
\caption{Summary of the exchange offers between \xrp and USD created in the \nw during 
 October 2016.\label{table:offers-oct16}}
\begin{tabular}{c c | c c | c}
Pay Val & Pay Cur & Get Val & Get Curr & Ratio\\
\hline
$1062738.51$ & XRP & $17009.50$ & USD & $62.48$ to $1$\\
$59678.62$ & USD & $33194.62$ & XRP & $1.78$ to $1$\\
\hline
\end{tabular}
\end{table}
\fi 

The results presented above 
describe the origin of a small fraction of the credit stuck on 
\links with \payroutes. We repeated the same experiment over the complete 
\nw lifetime and observed similar patterns. First, the \links with stuck credit 
are involved in a total of $278$ transactions where other \accs in the \nw are 
sending credit to victim \accs at a favorable rate: The receiver gets more credit 
than actually sent by the sender. Those transactions account for around 
 $158,000$ USD. Second, the highest amount of credit is received as a result of 
 circular \txs that use advantageous offers. 
 In particular, we find that \links with stuck credit are involved in a total 
 of $16,469$ \txs where they gained more than $63$M USD over the complete 
 \nw lifetime.

\paragraph{Countermeasures} 
Although \accs with stuck credit at \payroutes obtained  considerable revenue, a broader perspective
reveals that it was a risky operation. 
For instance, it is possible to check the exchange rates available in the \nw at October 2016 
to determine how likely it is to get the USD credit back. In particular, we observe that 
although \accs in \stuckacc{Oct}{16} managed to get ``cheap'' USDs, the market values 
were not favorable to 
get them back: New exchange offers created in the \nw in October 2016 (as shown 
in~\cref{table:offers-oct16}) demonstrate this.




\section{Stale Offers in the \nw}
\label{sec:stale}

Exchange offers and rippling are the key operations that enable path-based transactions. The previous two sections investigated the security of rippling, so we now investigate the safety of exchange offers, which are set by the owners of \accs at their own discretion. 
Naturally, proposed offers should match those of the corresponding currencies in the real world or 
even be in favor of market makers so that they get credit for their exchange services. 
Otherwise, cunning users can leverage stale offers to gain credit, while market makers may go bankrupt. This would adversely impact the liquidity and availability of the \nw.

\paragraph{Goal} In this experiment, we aim to determine whether there are stale offers in the 
\nw and, if so, study to what extent devilry users have taken advantage of them. 

\paragraph{Methodology}
We search for sudden changes in a currency's 
market capitalization.  
We observed several such changes. We first examine a spike in the price of XRP in 
late 2013:  
during a period of ten days (Nov 20th--30th, 2013), the price of $1$ \xrp with respect to BTC increased by $380\%$, i.e., $1$ \xrp was exchanged at 
 $0.00001$ BTC at the beginning of the period but within a week, $1$ \xrp 
 was exchanged at $0.000038$ BTC. Given that, we extract from \txdataset the transactions that occurred during this ten-day period, obtaining a total of $1,932$ transactions. We prune this dataset by 
 considering only cross-currency transactions that transfer \xrp for BTC or vice versa. We obtain 
 a total of $112$ transactions.  
 


We compare the exchange rate between \xrp and BTC used 
in each transaction to the exchange rate 
in the real world at the same time, as shown in~\cref{fig:TakerPaysBTC}. 
In both (top and bottom) figures, a purple point represents the exchange rate in a Ripple transaction while the corresponding green point 
denotes the exchange rate in the real world at the same time. For both graphs, if the purple point is higher than the green point (Ripple's offer is more 
expensive than the real world offer), the market maker made money. In contrast, if the 
purple point is below the green point, the user who conducted the transaction gained 
credit. 


\paragraph{Results} We analyzed  
the transactions in which a sender gained credit by exploiting stale offers. 
We make two observations. First, users could have gained up to around 
$250,000$ USD by fully exploiting \xrp/BTC stale offers during the specified period. In other words, 
market makers put at risk around $250,000$ USD due to stale offers. 
Second, $24$ different \accs made a monetary benefit of at 
least $7,500$ USD by exploiting \xrp/BTC stale offers (and other 
offers available in the network at that time). Here, we 
calculate the USD value by converting the BTC and \xrp to their real world exchange rates 
at the corresponding times. In summary,  
even in the nascent stages of the \nw, when the transaction 
volume was considerably low, 
stale offers risked significant loss of credit by market makers.

To confirm these results, we explored another, more recent, substantial change in a currency exchange rate. We found a sudden increase in the price of BTC compared to \xrp 
in 2017, concretely during the period July 16th  -- August 16th: The value of $1$ BTC
 went from $11,713$ \xrp
to $25,735$ \xrp, that is, an increase of $120\%$. 
As before, we extracted the transactions during that period of time and compared the 
exchange rates of \xrp from/to BTC in the \nw and in the real world. 
We observe that market makers put at risk around $500,000$ USD due to stale 
offers exchanging \xrp to BTC and vice versa. 
Moreover, we observe that $84$ \accs exploited these stale offers (and 
possibly other offers) to gain more than $4.5$M USD. 
These results confirm that stale offers continue to be a risk for market makers. In fact, the effect of stale offers is now amplified given the growth of the \nw and \txs. 

\paragraph{Countermeasures} A market maker  can update a previously offered exchange rate at any time. 
Therefore, a market maker should continuously monitor the price for the currencies involved in its offers and 
promptly update its Ripple offers when a sudden change occurs in the real world. 
The gaps between exchange rates in the \nw and real world are thereby reduced, and with them, the windows for cunning users 
to gain credit. 

\iffullversion
\begin{figure}[tb]
\centering
\includegraphics[width=0.49\columnwidth]{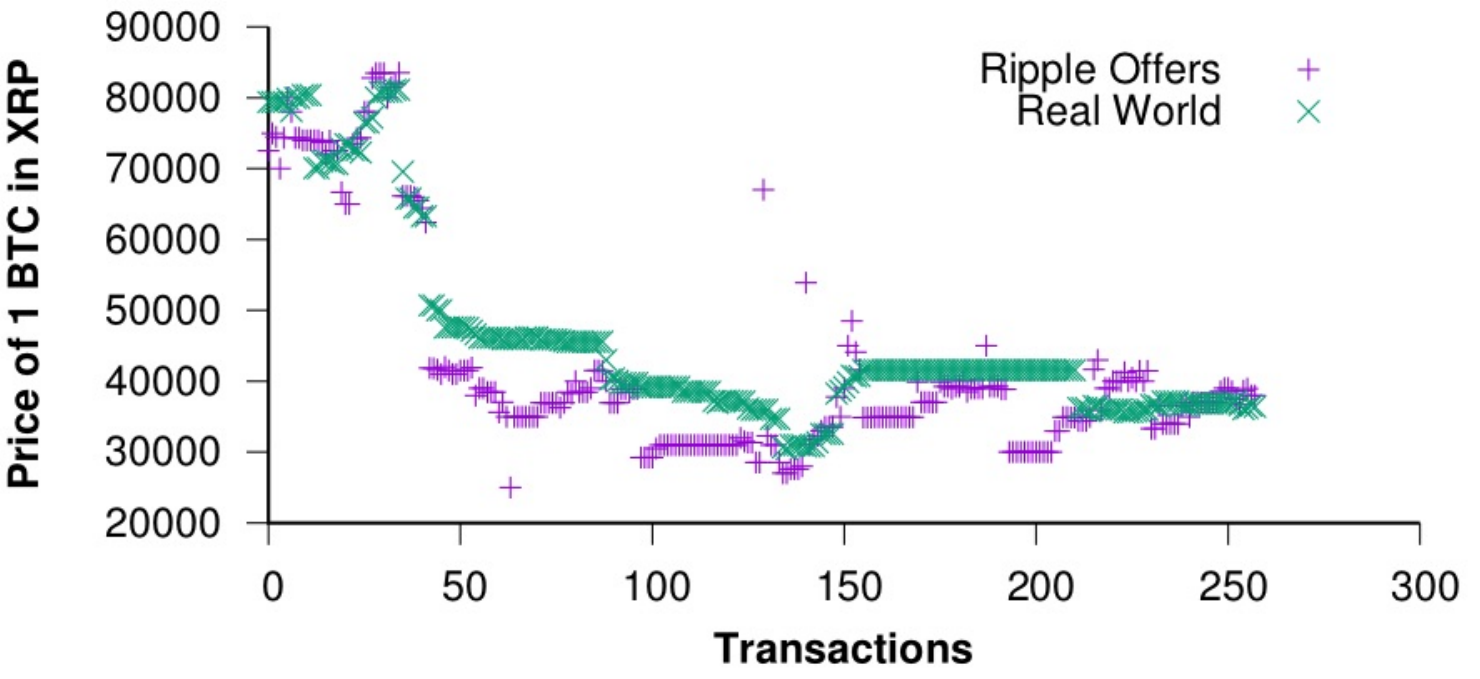}
\hfill
\includegraphics[width=0.49\columnwidth]{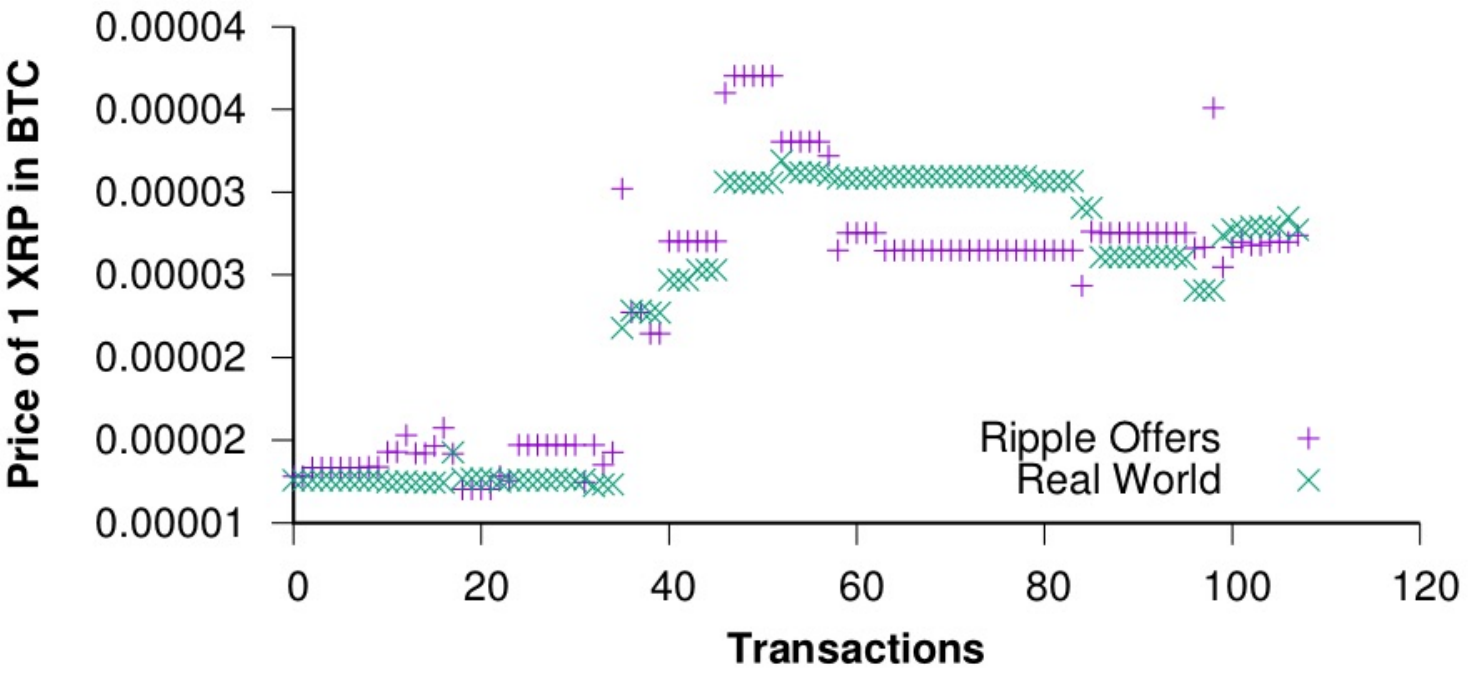}
\caption{Market maker accepts XRP and pays BTC (left); market maker accepts BTC and pays XRP (right). 
If the purple point (offer in Ripple) is below the green point (offer in real world), 
the transacting user gained credit. Otherwise, the market maker gained credit. These \txs took place between November 20th and 30th, 2013. \label{fig:TakerPaysBTC}}
\end{figure}
\else
\begin{figure}[tb]
\centering
\includegraphics[width=0.98\columnwidth]{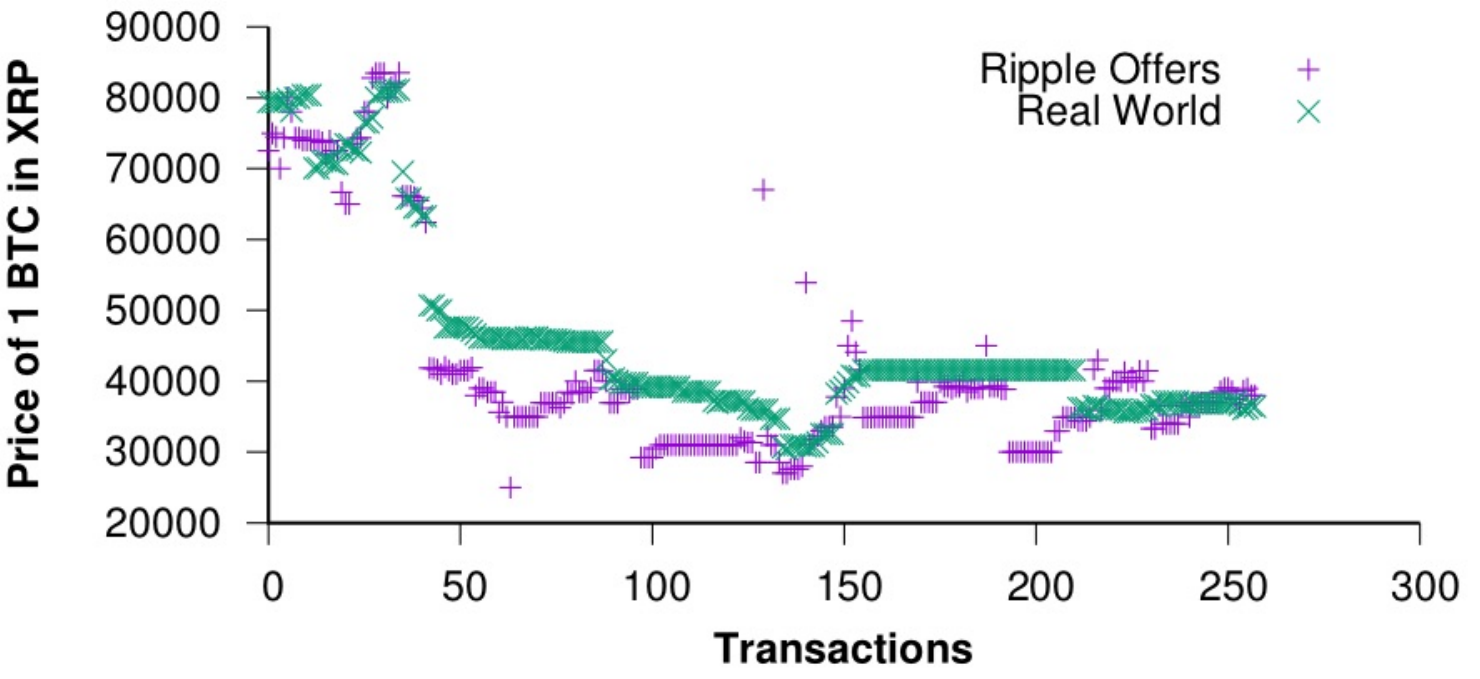}
\includegraphics[width=0.98\columnwidth]{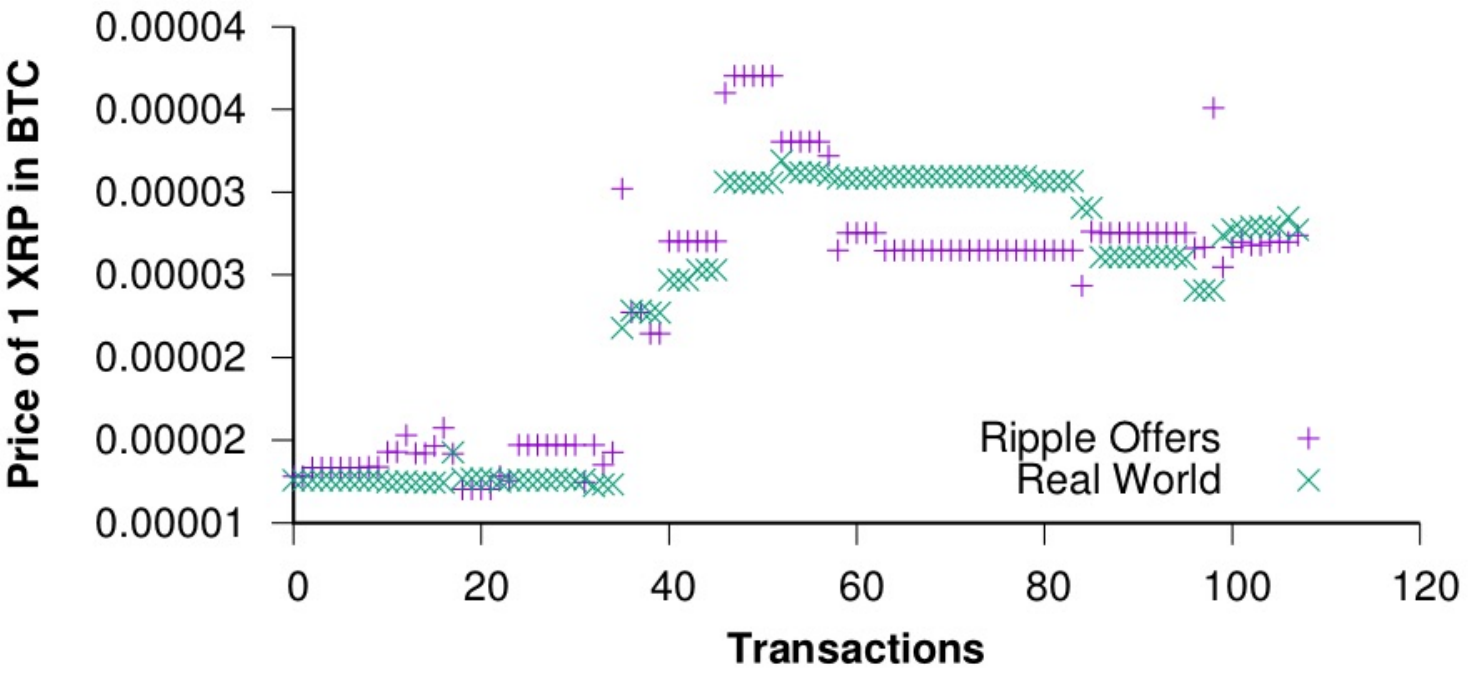}
\caption{Market maker accepts XRP and pays BTC (top); market maker accepts BTC and pays XRP (bottom). 
If the purple point (offer in Ripple) is below the green point (offer in real world), 
the transacting user gained credit. Otherwise, the market maker gained credit. These \txs took place between November 20th and 30th, 2013. \label{fig:TakerPaysBTC}}
\end{figure}
\fi



\section{Related Work}
\label{sec:relatedwork}

Some research work~\cite{Meiklejohn2013,Bonneau15,Ron2013,Barber2012,DonetDonet2014} 
has studied Bitcoin and other cryptocurrencies. Although it is possible 
to extract lessons from that work, the conceptual differences between 
cryptocurrencies such as Bitcoin and the \nw mandate a dedicated look.

There is limited work studying path-based settlement networks.
Moreno-Sanchez et al.~\cite{linkingwallets} present the first detailed study of the 
Ripple network. In particular, they identify the privacy breaches of the publicly available 
ledger. Their study links wallets that belong to the same user and 
deanonymizes the transactions associated with the main gateways. 
Their follow-up work~\cite{pathshuffle} presents a path-mixing 
protocol to allow anonymous transactions in the Ripple network, thereby mitigating privacy breaches. 

Di Luzio et al.~\cite{icdcs2017ripple} consider two aspects of the Ripple network. They study the evolution of the 
amount and behavior of participants in the consensus protocol used to add transactions to the ledger during the first three years of 
the Ripple network. They also propose a novel technique to deanonymize the 
transactions of a given user, leveraging side-channel information (e.g., the 
amount of a recent transaction performed by the victim). 

Armknecht et al.~\cite{Armknecht2015} present an overview of the Ripple network 
and give statistics about the number of transactions, and types of transactions and exchanges. The work is  
limited to the first two years of operation of the Ripple network. The work also demonstrates 
the conditions under which the Ripple consensus protocol fails, leading to a situation where 
the Ripple ledger might be forked. 
 
In summary, related work studies two dimensions of the Ripple network: 
privacy and consensus. Some work such as~\cite{icdcs2017ripple,Armknecht2015}
also computes statistics about the network structure during the first few years of the 
network lifetime.  In this work,  
we consider the consensus protocol and privacy as interesting but orthogonal 
dimensions to be studied, and instead focus on the evolution of the Ripple network and its vulnerabilities during the \emph{complete} network 
lifetime through August 2017. We thoroughly study several \emph{security} vulnerabilities and their 
implications on the Ripple network. 


\section{Concluding Remarks}
\label{sec:conclusions}

The \nw has been gaining momentum, with substantial growth in the number of \accs and \links in 2017. 
Yet, new \accs create \links with only few other 
key \accs, primarily gateways. This makes the \nw slow-mixing, with \accs grouped 
in demarcated communities. The users tend to stay bound to the same geographical community, elevating the importance of gateways in shaping the \nw. The core of the network composed of around $65,000$ \accs 
provides sufficient liquidity for the remaining \accs. 

The key operations in the \nw such as rippling and exchange 
offers pose important security challenges. Although the core of the \nw is resilient, a large number of users may be vulnerable to undesirable shift of credit among their \links. 
Thanks to the locality of communities, there is hope to tackle these vulnerabilities through geo-political forces. Further, users can be affected by the disruption of a handful of nodes, 
as demonstrated in the case of \payroutes, and hence are advised to add credit links. 
Last but not least, due to the importance of exchange offers in the current \nw, market makers are 
advised to periodically update their offers according to the real-world exchange rates, 
as they otherwise risk  several hundreds of thousands of dollars. 

Although this work focuses on the \nw, we believe that our findings are relevant to other emerging credit networks (e.g., Stellar~\cite{stellar-website}) 
and credit network-based systems~\cite{bazaar,ostra,iolaus,privpay,silentwhispers} that leverage similar design principles and may therefore present similar structural patterns and vulnerabilities.

\paragraph{Acknowledgments} 
We thank the Ripple forum members and the anonymous reviewers for their feedback.
This work is partially supported by an Intel/CERIAS RA and by the National Science Foundation under grants CNS-1319924 and CNS-1717493.

\balance
\bibliography{main}
\iffullversion
\bibliographystyle{acm}
\else 
\bibliographystyle{ACM-Reference-Format} 
\fi

\end{document}